%% file: main.tex
\documentclass[sigconf, nonacm]{acmart}

\newcommand{\showcomments}{0}

\newcommand\vldbpagestyle{plain} 

\usepackage{tikz}
\usepackage{xcolor}
\usepackage{tabularx} % Needed by std_commands
\usepackage{enumitem}

\usepackage{multicol}
\usepackage{subfig}
\usepackage{graphicx}

\usepackage{graphicx}
\usepackage{pifont}% http://ctan.org/pkg/pifont

\usepackage[framemethod=TikZ]{mdframed}

\definecolor{lightblue}{rgb}{0.905, 0.949, 1}
\definecolor{lightgreen}{rgb}{0.949, 1, 0.905}
\definecolor{darkgreen}{rgb}{0, 0.607, 0}
\definecolor{lightviolet}{rgb}{0.894, 0.886, 0.969}
\definecolor{darkviolet}{rgb}{0.423, 0.192, 0.612}
\definecolor{lightpink}{rgb}{0.969, 0.914, 0.953}
\definecolor{darkpink}{rgb}{0.827, 0.071, 0.773}

\newcommand*\circled[1]{%
  \begin{tikzpicture}[baseline=(C.base)]
    \node[draw=blue, text=blue, fill=lightblue, circle, line width=1pt, inner sep=1pt](C){\footnotesize{#1}};
  \end{tikzpicture}
}

\newcommand*\gcircled[1]{%
  \begin{tikzpicture}[baseline=(C.base)]
    \node[draw=darkgreen, text=darkgreen, fill=lightgreen, circle, line width=1pt, inner sep=1pt](C){\footnotesize{#1}};
  \end{tikzpicture}
}

\usepackage{dblfloatfix}

\begin{document}

\input{std_commands}
\setcounter{page}{1}

\title{Dynamic read \& write optimization with \TurtleKV}

\newcommand{\Tufts}{Tufts University}
\newcommand{\MathWorks}{MathWorks}

%%
%% The "author" command and its associated commands are used to define the authors and their affiliations.
\author{Tony Astolfi}
\affiliation{%
  \institution{\MathWorks}
}
\affiliation{%
  \institution{\Tufts}
}
%\email{tastolfi@gmail.com}

\author{Vidya Silai}
\affiliation{%
  \institution{\MathWorks}
}

\author{Darby Huye}
\affiliation{%
  \institution{\Tufts}
}

\author{Lan Liu}
\affiliation{%
  \institution{\Tufts}
}

\author{Raja R. Sambasivan}
\affiliation{%
  \institution{\Tufts}
}

\author{Johes Bater}
\affiliation{%
  \institution{\Tufts}
}

\input{sections/0-abstract}

\maketitle

%%% do not modify the following VLDB block %%
%%% VLDB block start %%%
\pagestyle{\vldbpagestyle}

\input{sections/1-introduction.tex}

\input{sections/2-bg-key-ideas.tex}

\input{sections/3-turtle-trees.tex}

\input{sections/4-turtle-kv.tex}

\input{sections/5-evaluation.tex}

\input{sections/6-related-work.tex}
\input{sections/7-conclusion.tex}

\newpage
\balance
\bibliographystyle{ACM-Reference-Format}
\bibliography{references.bib}
\nobalance

\end{document}

%% file: std_commands.tex
%-- place any standard commands/environments here to get included in
%-- documents.  When you include this file, you should do it before
%-- the \begin{document} tag.

%%%%%%%%%%%%%%%%%%%%%%%%%%%%%%%%%%%%%%%%%%%%%%%%%%%%%%%%%%%%%%%%%%%%%%
%-- CHANGES:
%-- 07/31/01 -jstrunk- Added command to set the paper margins.

%-- Provides fixed width font for commands and code snips.
\newcommand{\code}[1]{\texttt{\textbf{#1}}}

%-- Terms...  Use this to introduce a term in the paper.
\newcommand{\term}[1]{\emph{#1}}

%-- Provides stylization for e-mail addresses
% \newcommand{\email}[1]{\emph{(#1)}}

%-- Starts a minor section (puts the title inline w/ the text.
\newcommand{\minorsection}[1]{\textbf{#1}:}

%-- Jiri caption
\newcommand{\minicaption}[2]{\caption[#1]{\textbf{#1.} #2}}

%-- Units on numbers: 4KB -> \units{4}{KB}
\newcommand{\units}[2]{#1~#2}

%-- Commands...  i.e. WRITE commands.
\newcommand{\command}[1]{{\sc \MakeLowercase{#1}}}

%-- For notes about things that need to be fixed.
\newcommand{\fix}[1]{\marginpar{\LARGE\ensuremath{\bullet}}
    \MakeUppercase{\textbf{[#1]}}}
%-- For adding inline notes to a draft preceded by your initials
%-- E.g., \fixnote{JJW}{What the heck is a foobar?}
\newcommand{\fixnote}[2]{\marginpar{\LARGE\ensuremath{\bullet}}
    {\textbf{[#1:} \textit{#2\,}\textbf{]}}}

%-- Setting margins: \setmargins{left}{right}{top}{bottom}
\newcommand{\setmargins}[4]{
    % Calculations of top & bottom margins
    \setlength\topmargin{#3}
    \addtolength\topmargin{-.5in}  %-- seems like this should be 1, but .5
                                   %-- balances the text top to bottom
    \addtolength\topmargin{-\headheight}
    \addtolength\topmargin{-\headsep}
    \setlength\textheight{\paperheight}
    \addtolength\textheight{-#3}
    \addtolength\textheight{-#4}

    % Calculations of left & right margins
    \setlength\oddsidemargin{#1}
    \addtolength\oddsidemargin{-1in}
    \setlength\evensidemargin{\oddsidemargin}
    \setlength\textwidth{\paperwidth}
    \addtolength\textwidth{-#1}
    \addtolength\textwidth{-#2}
}

%-- For the tabularx environment... Using L, C, R as the column type
%-- will left, center, or right justify the text.
\newcolumntype{L}{X}
\newcolumntype{C}{>{\centering\arraybackslash}X}
\newcolumntype{R}{>{\raggedleft\arraybackslash}X}

%-- To comment out a swatch of text, use \omitit{blah blah blah}
\long\def\omitit#1{}

%-- Inline title; useful for sub-sub-sections in which you don't want a separate
%-- line for the title.
\newcommand{\inlinesection}[1]{\smallskip\noindent{\textbf{#1.}}}

\newenvironment{outlineenv}{\par\color{blue}}{\par}
\newenvironment{pagelenenv}{\par\color{red}}{\par}

\newcommand{\outline}[1]{\begin{outlineenv}#1\end{outlineenv}}
\newcommand{\pagelenblah}[1]{\begin{pagelenenv}#1\end{pagelenenv}}

\newcommand{\pagelen}[1]{
   \ifthenelse{\equal{\showcomments}{1}}{
     \pagelenblah{#1}}{}}

\newcommand{\outlinetext}[1]{
   \ifthenelse{\equal{\showcomments}{1}}{
     \outline{#1}}{}}

%-- For missing data that needs to be filled in
\newcounter{missingctr}
\newcommand{\missingval}[1]{\stepcounter{missingctr}\underline{\textbf{MISSING\_VAL~\arabic{missingctr}}}}

% Colorred number for inline items
\newcommand{\itm}[1]{\textcolor{purple}{\textbf{(#1)}}}

% Colored numer for inline items, emphasized
\newcommand{\iitm}[1]{\textbf{\textcolor{purple}{\emph{(#1)}}}}

% Command to contributoin names
\newcommand{\toolname}{TurtleKV}

%%%%%%% custom commands to TurtleKB paper

\newcommand{\Btree}{B-tree}
\newcommand{\Bwtree}{Bw-tree}
\newcommand{\Bftree}{Bf-tree}
\newcommand{\Bplustree}{B$^+$-tree}
\newcommand{\Bepsilontree}{B$^{\epsilon}$-tree}
\newcommand{\STBepsilontree}{STB$^{\epsilon}$-tree}
\newcommand{\Bepsilonplustree}{B$^{\epsilon+}$-tree}
\newcommand{\TurtleKV}{TurtleKV}
\newcommand{\TurtleTree}{TurtleTree}
\newcommand{\LSMtree}{LSM-tree}
\newcommand{\checkpointdistanceM}{\chi}
\newcommand{\checkpointdistance}{$\checkpointdistanceM$}

% --------- params
\newcommand{\MemTableSize}{$\overline{M_w}$}
\newcommand{\MemTableCount}{$|M_w|$}
\newcommand{\MemTableSizeM}{\overline{M_w}}
\newcommand{\MemTableCountM}{|M_w|}

\newcommand{\cmark}{\ding{51}}%
\newcommand{\xmark}{\ding{55}}%

%% file: sections/0-abstract.tex
%%
%% The abstract is a short summary of the work to be presented in the
%% article.
%-------------------------------------------------------------------------------
\begin{abstract}
\vspace{0.15cm}

High read and write performance is important for generic key-value
stores, which are foundational to modern applications and databases.
Yet, achieving high performance for mixed and dynamic workloads is
challenging due to fundamental trade-offs between memory use and I/O
for retrieval and updates. Past work emphasizes the trade-off between read-
and write-optimization as expressed through primary data structure, in
combination with read-memory trade-off mechanisms like caching and
filtering.  This raises re-tuning costs as optimal trade-off targets
change, due to restructuring of stored data.  We show that
write-memory trade-off mechanisms are under-developed in current
designs, and propose a new approach to \textit{dynamic key-value store
  optimization} using a novel read-/write-balanced on-disk structure,
the \TurtleTree{}, and flexible read-memory \& write-memory tuning
knobs.  We describe how the design of \TurtleKV{}, our prototype,
avoids in-memory bottlenecks to achieve high performance across a wide
range of tuning parameters.  When evaluated using YCSB, \TurtleKV{}
matches state-of-the-art SplinterDB for inserts, and is
5$\times{}$/12$\times$ faster than RockDB/WiredTiger.  In mixed
workloads, \TurtleKV{} is 16-25\% faster than SplinterDB, >4$\times$
RocksDB, and 3-6$\times$ WiredTiger.  \TurtleKV{} is 2-9$\times$
faster than the others for point-query workloads, and has the best
scan performance of the write-optimized systems tested.
%-------------------------------------------------------------------------------
\end{abstract}

%% file: sections/1-introduction.tex
\section{Introduction}
\label{sec:intro}

Today the builders of complex, cloud-based applications and systems
have many different key-value stores to choose from, each with their
own biases and trade-offs (e.g., read-optimized vs write-optimized;
online vs analytic; in- vs out-of-memory; ordered vs hashed;
domain-specific vs general-purpose; etc.) \cite{
  jermaineNovelIndexSupporting1999, chaBlinkHashAdaptive2023,
  giladEvenDBOptimizingKeyvalue2020, pelkonenGorillaFastScalable2015,
  ashkianiGPULSMDynamic2018, wangLavaStoreByteDancesPurposeBuilt2024,
  sollezaMachPluggableMetrics2022,
  renSlimDBSpaceefficientKeyvalue2017a}.  These are frequently
difficult to tune to accommodate evolving requirements after a system
is built \cite{RocksDBTuningGuide,
  caoCharacterizingModelingBenchmarking2020, SmallDatumImpact2022,
  jagadeesanWorstCaseAverageCaseAnalysis2020, RevisitingB+treeVs2022,
  nguyenCharacteristicsMongoDBTrimbased2016,
  andreoliInducingHugeTail2023,
  fedorovaPerformanceComprehensionWiredTiger2018}.  As a consequence,
we see a proliferation of multi-stores with complex and expensive
logic for synchronization, data transfer, and invariant maintenance
\cite{etl_pipeline, kumarValuePropositionETL2019,
  mukherjeeComparativeReviewData2017, el-sappaghProposedModelData2011,
  limSILTMemoryefficientHighperformance2011}. Engineering effort must be
spent multiple times for development, debugging, optimization, and
maintenance of each storage engine.

One would not be without reason believing that a multitude of storage
engines is inevitable.  The RUM Conjecture
\cite{athanassoulisDesigningAccessMethods2016} posits a fundamental,
three-way trade-off in the design of data storage/access methods
between the resource overhead of reads (R), updates/writes (U), and
memory/storage (M).  If correct, this might seem to imply that the
divergence of requirements for different workloads and applications
means key-value stores inhabit a problem space where ``one size'' does
not, and cannot, fit all \cite{idreosKeyValueStorageEngines2020}.  Our
work assumes the correctness of the RUM Conjecture as its starting
point, but seeks to challenge this implication and propose an
alternative approach to key-value storage engine design in the face of
inherent trade-offs.

The notion of a three-way trade-off in the RUM Conjecture means that 
at least two trade-off selection mechanisms, or \textit{tuning knobs}, must be
utilized to fully cover the space of optimal possibilities;
e.g. reads-vs-writes (RW), reads-vs-memory (RM), or writes-vs-memory
(WM).  Our key insight is that these basic trade-offs have important
differences, and that these differences make some inherently better
suited for use as tuning knobs.  For example, if a suboptimal
tuning point is selected for the read-memory trade-off, it is likely
easy to correct because of the mechanisms involved: simply modify the
cache size to decrease either the I/O cost of queries or the memory
footprint of the system at the expense of the other.

In contrast, if a
system is improperly tuned with respect to the read-write trade-off,
things aren't so simple.  This is because read-write tuning is largely
accomplished via selection of primary data structure, which affects
implementation (code) and the invariants of stored data, both of which
are expensive to modify post-facto.  While some dictionary data
structures do support parameter-based read-write trade-off tuning
(e.g., compaction policy in \LSMtree{}s with adjustable fanout
\cite{oneilLogstructuredMergetreeLSMtree1996,
  luoLSMbasedStorageTechniques2020}), when analyzed closely and
evaluated empirically, we find the mechanisms in common use to be
sub-optimal.  In any case, even if the primary data structure of a
system does support smooth, optimal tuning between read and write
optimization (e.g, \Bepsilontree{}s' $\epsilon$ parameter
\cite{brodalLowerBoundsExternal2003,
  benderIntroductionBetreesWriteOptimization2015}), bias is still
encoded at the storage level, making re-tuning expensive.

Since write-optimization is very important to certain workloads
\cite{caoCharacterizingModelingBenchmarking2020}, write-memory (WM)
tuning must be supported as a first-class feature if we wish to avoid
the downsides of RW tuning bias in stored data structures.  To do this
it is clearly necessary, though insufficient, to simply expose a
tuning knob.  The choice of data structure and the overall design of
the system must work together to maximize the effectiveness of this
trade-off.  A good example of how data structures can anticipate a
resource trade-off (RM) in a system design context is the pervasive
use of \Bplustree{}s in preference to \Btree{}s
\cite{comerUbiquitousBTree1979,
  bayerOrganizationMaintenanceLarge1970}.  By storing all key-value
records at the bottom level of the tree, \Bplustree{}s offer increased
branching factor and the possibility for all interior nodes to be
cached in memory, reducing read amplification to $O(1)$. A key idea of
the work presented here is to apply a similar principle in data
structure selection to enhance the effectiveness of the write-memory
(WM) trade-off, elevating it to first-class status in key-value system
design.

We present \TurtleKV{}, a proof-of-concept for a new approach to
key-value storage oriented around RM and WM tuning knobs which
dynamically navigate the RUM trade-off space without having to change
the structure of stored data. \TurtleKV{} is built around a novel data
structure, the \TurtleTree{}, which offers two important properties:
1. it achieves a neutral, balanced compromise between the extremes of
read- and write-optimized external memory dictionaries, and 2. it
affords a smooth trade-off between memory usage and write
amplification which is simultaneously efficient, scale-independent (in
terms of data size $N$), conceptually simple, and wide-ranged in
effect.  \TurtleKV{} implements a system architecture which is similar
at a high level to conventional LSM-based key-value stores, with the
addition of several crucial optimizations to eliminate in-memory
bottlenecks so that reductions in write-amplification translate to
increased throughput.  Because \TurtleKV{} starts from a fundamentally
stronger premise, it avoids many of the complex workarounds and
features which tend to creep into modern key-value stores in order to
try to mitigate their intrinsic limitations.  As a result, we show
that \TurtleKV{} outperforms today's key-value engines, while offering
superior tunability, with a modest code base of similar or lesser
complexity.

In this work we demonstrate that write-vs-memory tuning mechanisms are
often not given first-class status in the design of current key-value
stores ($\S$\ref{systemTuningConcerns}).  For example, while RocksDB
\cite{RocksDBPersistentKeyvalue}, one of the most popular key-value
storage engines today, supports runtime adjustment of write buffer
size \cite{RocksDBTuningGuide}, and our experiments show this
parameter is highly effective at spending memory to reduce write
amplification, doing so yields very little benefit in terms of
increased end-to-end performance.  WiredTiger
\cite{WiredTigerWiredTigerDeveloper}, another popular storage engine
on the opposite end of the read-write spectrum, also supports WM
tuning, and even demonstrates end-to-end benefit in terms of increased
insert throughput over wide range of tuning points.  However,
WiredTiger's \Bplustree{} based design offers little benefit at
low-to-moderate memory allocations due to the inherent limitations of
the data structure \cite{idreosDesignContinuumsPath2019}.  Finally, we
find that SplinterDB \cite{conwaySplinterDBClosingBandwidth2020,
  conwaySplinterDBMapletsImproving2023,SplinterDBb} which is optimized
for high concurrency and fast storage devices, is able to take
advantage of higher memory allocations to reduce write amplification
and average insertion latency for randomized insertion workloads.
However, since SplinterDB does not offer any way to express a
preference for optimization target (i.e., use cache memory to optimize
writes vs reads), in mixed workloads the optimization effect of higher
cache size may become unpredictable, as the dueling concerns of
updates and queries vie for limited memory resources.

The contributions of this work are:\begin{enumerate}[leftmargin=*]
%\vspace{-0.1cm}

\item An analysis of how current key-value stores and their data
  structures fall short of fully exploiting the potential of the RUM
  space, especially with respect to the \textit{write-memory
    trade-off}
\item A novel external memory dictionary type, the
  \textit{\TurtleTree{}}, within a previously unidentified class of
  data structures, \textit{\Bepsilonplustree{}s}, which combines ideas
  from both read- and write-optimized dictionary types
\item A key-value storage engine design, \textit{\TurtleKV{}}, which
  uses \TurtleTree{}s to offer simple and smooth trade-off tuning for
  a wide range of resource allocations
\item An experimental evaluation of \TurtleKV{} as compared to SOTA
  key-value stores WiredTiger, RocksDB, and SplinterDB
\end{enumerate}

%% file: sections/2-bg-key-ideas.tex
\section{Read, Update, Memory: Pick Two}\label{section:background_key_ideas}

\input{figures/fig_chivslazytuning}

The RUM Conjecture \cite{athanassoulisDesigningAccessMethods2016}
states that if one holds any of the three resource overheads (read,
update/write, and memory/space) constant, the remaining two are placed
in a zero-sum trade-off relationship.  Geometrically, this can be
visualized, as in Figure \ref{chiVsLazyTuning}, by a three dimensional
resource space, where points represent the actual resource cost of a
concrete data access method with respect to a particular workload.
The notion of a three-way trade-off, in this conceptualization, is
represented by a two dimensional surface or manifold embedded within
the larger-dimensional space (dark gray).  Fixed resource commitments
can be thought of as two dimensional slices of this surface orthogonal
to the axis of the fixed resource, with the resulting intersection
defining the trade-off curve between the remaining two resources.

Navigating the optimal two-dimensional subset of the RUM space
requires the use of exactly two mechanisms, or tuning knobs, which
serve as a coordinate system within the sub-space.  In this section we
analyze our problem space by considering three tuning knob candidates:
the read-write trade-off (RW), read-memory trade-off (RM), and
write-memory trade-off (WM).  These are explored through the
perspectives of data structure ($\S$\ref{readWriteDataStructures}) and
system design ($\S$\ref{systemTuningConcerns}).  We conclude with a
sketch of the key ideas behind \TurtleKV{}
($\S$\ref{keyIdeasSection}).

%\vspace{-0.5cm}
\subsection{Data Structures: Trade-off Tuning}\label{section:bg:data_structures}\label{readWriteDataStructures}

 \begin{figure*} \centering \includegraphics[width=1 \textwidth]
 {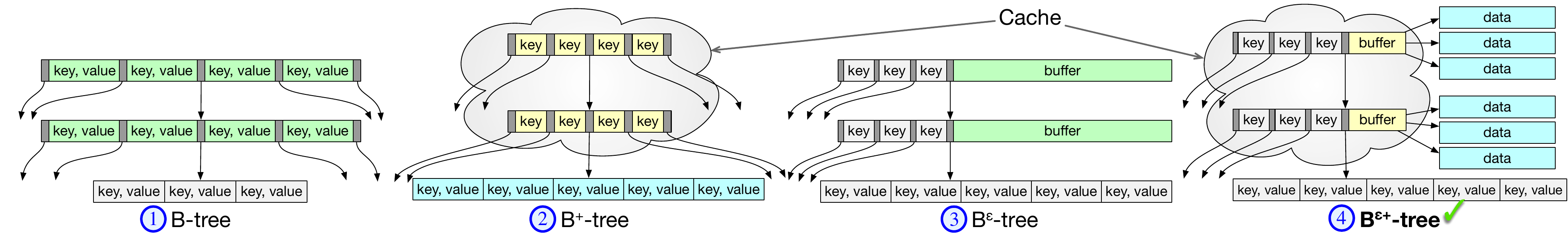} 
 \vspace{-0.5cm}
 \caption
 {B$^\epsilon$-tree $\rightarrow$ B$^{\epsilon+}$-tree structural
 transformation}\label{bEpsilonPlusTransformation}
 \vspace{-0.25cm}
\end{figure*}

The most consequential design decision for a key-value storage engine
is its choice of primary data structure.  Here we review the major
categories of external memory dictionary, grouped by their RW
trade-off optimization target (read-optimized, write-optimized, and
hybrid/balanced).  We highlight the essential insights of each data
structure, how each approaches trade-off tuning, and their inherent
limitations.

\vspace{-0.15cm}
\subsubsection{Read-Optimized (B-trees).}\label{section:bg:data_structures:read_optimized}
\Btree{}s\cite{comerUbiquitousBTree1979} provide optimal~\cite
{brodalLowerBoundsExternal2003, yiDynamicIndexabilityOptimality2012} read performance and can
utilize memory to reduce the number of reads required to process
queries by caching pages containing tree-nodes (RM trade-off).
\Bplustree{}s, the most commonly implemented B-tree variant, exploit
this mechanism by placing key-value data in the leaves, allowing all
but the bottom layer of the tree to be cached in main memory so that
queries require only a single read.

State-of-the-art approaches to reducing the write cost of updates in
B-trees employ the use of a write-ahead log (WAL) to ensure durability
of updates, plus a scheme for applying updates to the tree in memory
(WM trade-off). We refer to these as \textit{Log Structured B-trees}.
Their goal is to defer rewriting pages containing updated keys, in
hopes of collecting enough updates per page to amortize I/O.
\Bwtree{}s~\cite{levandoskiBwTreeBtreeNew2013} do this through the use of delta records
applied in-memory to the main tree using lock-/latch-free techniques.
The closely related \Bftree{}~\cite{bftree} uses an alternate strategy
of caching and updating sub-page slices (\textit{mini-pages}) instead
of full pages, which has additional benefits in terms of lowering the
memory cost of caching ``hot'' records.

\Bwtree{}s and \Bftree{}s achieve good write efficiency either when memory
is plentiful or when updates are clustered in a few sub-ranges of the
key space.  In the general case, however, Log Structured
B-trees force a sub-optimal trade-off between memory and
  write amplification, especially as the total data size $N$
   grows with respect to available memory $M$
\cite{idreosDesignContinuumsPath2019}.

\vspace{-0.15cm}
\subsubsection{Write-Optimized (LSM-trees)}\label{section:bg:data_structures:write_optimized}

Log-Structured Merge (LSM) trees
\cite{oneilLogstructuredMergetreeLSMtree1996} are a popular
write-optimized alternative to \Btree{}s.  By deferring the merge (or
\textit{compaction}) of new updates with existing data, they maximize
the amount of effective work achieved by each write operation.  This
results in a dramatic reduction in write amplification, but presents
two challenges: first, it increases read costs (and storage footprint)
by introducing the possibility that updates for a given key may be
stored in multiple locations.  Second, since the increasingly high
bandwidth of modern storage devices favors performing merges in large
sequential runs to maximize I/O utilization, update costs in
\LSMtree{}s tend to be \textit{amortized} costs; i.e. tail latency
can be quite high.  This motivates the need for effective
de-amortized, or incremental, approaches to compaction.
 
Page caching is less effective as an RM tuning mechanism for
\LSMtree{}s than \Btree{}s because there is potentially much more data
to cache. In addition to caching, two primary tuning mechanisms are
employed by \LSMtree{} systems: compaction policy and approximate
membership query (AMQ) filtering.  Compaction policies divide into
\textit{greedy} (level-tiering) and \textit{lazy} (size-tiering).  In
a level-tiered \LSMtree{}, write cost is increased by repeatedly
merging data at the same level (up to $F$ times) to reduce future read
cost.  In a size-tiered \LSMtree{}, a tunable number $T$ of key
duplicates are allowed to co-exist at a given level before their
eventual merge into a larger level.  Compaction policy thus alters the
invariants of stored data to achieve a RW trade-off mechanism.

Approximate member query (AMQ) filters, e.g. Bloom Filters
\cite{bloomSpaceTimeTradeoffs1970}, are an RM trade-off mechanism that
reduces the cost of point queries by having a dedicated filter (or part of a
shared filter \cite {conwaySplinterDBMapletsImproving2023}) devoted to
each location where a key might reside.  AMQ filters offer a tuning
parameter $\alpha$\footnote{Also frequently named as $\epsilon$}, the
false-positive rate, which trades size against accuracy.\footnote{A
filter must always answer accurately if a query key is not in its set,
but it may give a false positive result at rate $\alpha$.  For lower
$\alpha$, more bits per key are required.}  To answer a point query for key $k$, the filters for all
locations $S$ where $k$ might reside are consulted before performing
any I/O.  This rules out, with probability $\prod_{i \in
  S}{\left(1-\alpha_i\right)}$, locations where the query will fail,
preventing excess reads.

If filters are used to maintain a target read cost in a size-tiered
\LSMtree{} (with $T$ tiers per level), the total size of the filters
must scale in proportion to $T$, since the total number of
false-positives for a query is the sum of false-positives for all
filters that pertain to a given key
\cite{dayanOptimalBloomFilters2018a}.  As $T$ increases, read
amplification increases while write amplification goes down; as filter
bits-per-key increases, memory usage rises while read amplification
falls due to lower false-positive rate $\alpha$.  It is theoretically
possible therefore to coordinate the motion of these two parameters
($T,\alpha$) to target a fixed level of read amplification, while
writes are effectively traded against memory (WM) using filter
size-vs-accuracy as an intermediary. We refer to this strategy as
\textit{filter arbitrage}.

\vspace{-0.15cm}
\subsubsection{Hybrid-Optimized (B$^{\epsilon}$-trees)}
\label{section:bg:data_structures:hybrid_optimized}
B$^{\epsilon}$-trees
\cite
 {benderIntroductionBetreesWriteOptimization2015,
 brodalLowerBoundsExternal2003} are a tunable external memory
 dictionary, able to trade reads against writes.  The basic B-tree
 structure is used as the foundation for a B$^{\epsilon}$-tree.  The
 $\epsilon$ parameter varies the portion of a node page used for
 pivots, which optimize for reading, and update buffering, which
 optimizes for writing.  At either extreme of their tuning range
 $\epsilon \in {0..1}$, B$^{\epsilon}$-trees simply degrade into the
 equivalent of a B-tree or LSM-tree.  At $\epsilon = \frac{1}
 {2}$, the branching factor is $\sqrt{B}$, making tree height twice
 that of an equivalent B-tree.  In this configuration, updates can
 be
\textit{flushed} down one level at a time from the root in batches of
 size $\Theta(\sqrt{B})$, dramatically lowering the worst-case write
 cost and providing a good balanced tuning point.

One disadvantage of traditional B$^\epsilon$-trees, where updates are
buffered inline within the node page (Figure
\ref{bEpsilonPlusTransformation}, \circled{3}), is that, for any given
 value of $\epsilon$, increasing the page size $B$ to optimize writes
 increases the I/O and/or memory cost of reading/caching a node.
 Similarly, decreasing $\epsilon$ to decrease the branching factor
 and increase the buffer size increases the height of the tree, which
 raises the cost of reading/caching a node-path to any given key.

Size-Tiered \Bepsilontree{}s (\STBepsilontree{}s), used by SplinterDB
\cite {conwaySplinterDBClosingBandwidth2020,
  conwaySplinterDBMapletsImproving2023}, are a recent re-imagining of
the B$^\epsilon$-tree designed to close the ``bandwidth gap'' between
the capabilities of fast NVMe SSDs and traditional key-value storage
engine designs.  The main structural modification in
STB$^\epsilon$-trees is to add a level of indirection between the tree
of interior node pages, or \textit{trunk}, and buffered data, or
\textit{branches} (Figure \ref{bEpsilonPlusTransformation},
\circled{3}$\rightarrow$\circled{4}).  This separation allows buffered updates to be flushed
in larger batches than a traditional \Bepsilontree{}, enabling high
I/O bandwidth utilization.  It also enables flushes to be decoupled
from compactions since only the branch references need to be changed,
enabling SplinterDB's novel \textit{flush-then-compact} policy.
Flush-then-compact has the added benefit of allowing updates with
skewed key distributions to be compacted at a lower level of the tree,
reducing write amplification for ``hot'' records.  However, due to their
lazy compaction policy, \STBepsilontree{}s must do more work to
perform scan queries, which hurts performance.

\subsection{System Design: Write-Memory Tuning}\label{systemTuningConcerns}

We now shift our attention from data structures to the concerns of
system design, to explore how well current key-value storage engines
support write-vs-memory (WM) trade-off tuning.  We present three case
studies of systems spanning the read-write (RW) spectrum covered in
$\S$\ref{readWriteDataStructures}: WiredTiger, RocksDB, and
SplinterDB.  All are high-performance key-value storage engines
optimized for concurrent workloads run on hardware with multiple
processor cores and solid-state storage devices.  RocksDB and
WiredTiger in particular are heavily relied upon in real-world
systems, and have been the beneficiaries of a large investment in
engineering effort.  The main differentiator in these systems is the
choice of primary data structure: WiredTiger uses
B$^+$-Trees\footnote{WiredTiger also supports LSM-based indexing, but
it is not the default.}, RocksDB is based on LSM-Trees, and SplinterDB
introduces the novel \STBepsilontree{}.

An investigation into the limitations of how each of these systems
support WM trade-off tuning elucidates some of the central issues and
provides motivation for the design of our prototype, \TurtleKV{}.
Specifically, we present three findings: first, a system's ability to
navigate the WM trade-off is highly dependent on the choice of primary
data structure; second, the ability to reduce write amplification does
not always translate into increased real-world performance; and third,
supporting I/O vs memory trade-offs in general does not imply the
ability to target specific tuning points in terms of memory-based
read- or write-optimization.

We summarize the key results of these case studies in Table \ref{table:db_comparison}.

\input{figures/table_db_comparison.tex}
%---------------------------------------------------------------------------------
\subsubsection{WiredTiger}\label{case_study:wiredtiger}

WiredTiger is the default storage engine for MongoDB. It is designed
to support high levels of concurrency for transactional key-value
workloads.  To accomplish this, it uses lock/latch-free algorithms and
multi-version concurrency control (avoiding update-in-place) to
implement its primary data structure, a \Bplustree{}.

Updates to the \Bplustree{} in WiredTiger take place in ``dirty''
buffers pinned to its in-memory page cache, where new versions of
existing pages (or entirely new pages) are built.  Dirty pages are
flushed to durable storage in response to a checkpoint request or when
their total size exceeds a runtime-configurable
threshold\footnote{\texttt{eviction\_dirty\_target} for background
threads, \texttt{eviction\_dirty\_trigger} for application/foreground
threads.}. Write-ahead logging can be enabled for durability of
record-level updates in between dirty page flush triggers.  Thus,
WiredTiger supports dynamic write-memory (WM) trade-off tuning through its
dirty page eviction (write-back) caching mechanism.

Figure \ref{figure:wiredtiger:wm} shows how WiredTiger responds to
increased memory allocation $M$ for dirty page buffers, running a workload
with uniformly distributed keys inserted in random order.  Here we see
that average insert latency tracks with write amplification through
the entire dynamic range of the resource trade-off, which is good.
Reductions in write amplification factor (WAF), however, are minimal
for 16MB $\le M \le$ 500MB; at higher memory allocations, WAF
reduction accelerates.  This is because, for random (non-skewed) key
distributions, the expected benefit of buffering in B$^+$-Trees is
related to the total dictionary size $N$.  If each page can hold $B$
items, then the expected write amplification $W_x$ for record $x$ is
$O(\max(1, \min(\frac{N}{M},B)) )$. Note, $W_x \in [1..B]$, since
each record must be written at least once ($1$), and in the worst case
it is re-written every time each other record in the same page is
written ($B$).

%---------------------------------------------------------------------------------
\subsubsection{RocksDB}\label{case_study:rocksdb}

RocksDB is used at Meta as the primary storage engine of ZippyDB (a
distributed key-value storage service developed by and used within
Meta), UDB (a MySQL storage layer for social graph data)
\cite{caoCharacterizingModelingBenchmarking2020}, and by countless other
open source projects and commercial systems.

RockDB's Tuning Guide \cite{RocksDBTuningGuide} presents a formidable
set of user-selectable policies and tuning knobs which optimize for
various use cases.  We identify two such knobs which appear to have
the greatest impact on the WM trade-off: the size \MemTableSize{} of
an in-memory table
(\textit{MemTable}) \footnote{\texttt{rocksdb.options.write\_buffer\_size}},
and the number of MemTables
\MemTableCount{}\footnote{\texttt{rocksdb.options.min\_write\_buffer\_number\_to\_merge},
\texttt{rocksdb.options.max\_write\_buffer\_number}} collected in
memory before being flushed to the first level of external storage
($L_0$).  We find that varying \MemTableSize{} with fixed
\MemTableCount{} is superior to varying \MemTableCount{} for fixed
size \MemTableSize{}, due to the degradation of query performance as
the number of MemTables increases; both approaches offer the same
benefit in write-optimization.

RocksDB's default compaction strategy (leveled) uses incremental
merging, in which levels are segmented into fixed-size sorted string
table (SSTable) files.  When the total size of all segments within a
level $L_i$ grows too large, an SSTable (of size \MemTableSize{})
within $L_i$ is selected to be \textit{compacted} with data in the
next larger/older level $L_{i+1}$.  The compaction merges the single
segment from $L_i$ with all segments in $L_{i+1}$ whose key range
overlaps, producing a new sequence of SSTable segments whose keys are
ordered and unique.  The total write amplification $W_x$ for a given
record $x$ can be calculated as the number of times $F$ (for
\textit{fanout}) the record is merged into a segment at a given level,
times the number of levels in the LSM\footnote{The reason there are
$\log_F({N/M})=\log_F{N}-\log_F{M}$ levels for data set size $N$ is
that the smallest level's size is always at least $M$; one can think
of the first $\log_F{M}$ merges of a record as happening in main
memory, inside a MemTable.}: $W_x = F \cdot
\log_{F}({N/\MemTableSizeM})$.  Increasing MemTable size $M$ therefore
spends memory to reduce write amplification, independent of total size
$N$: $\Delta W_x = -O(\log{\MemTableSizeM})$.

\begin{figure}
  \vspace{-0.4cm}
  \centering 
	\subfloat[\centering WiredTiger]{{  
	  \hspace{-0.25cm}\includegraphics[width=0.36 \columnwidth]{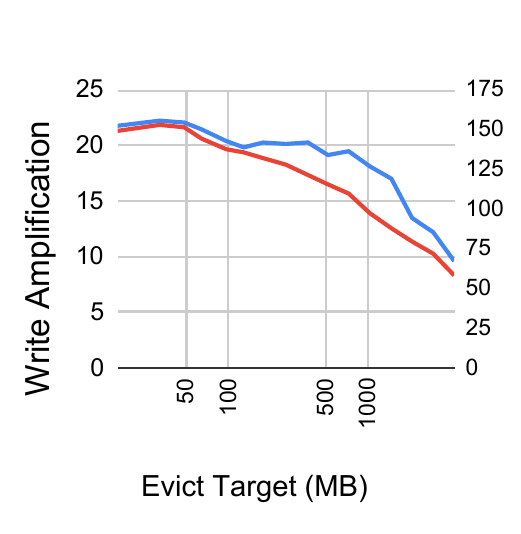}\label{figure:wiredtiger:wm}
	}}
	\subfloat[\centering RocksDB]{{  
	  \hspace{-0.21cm}\includegraphics[width=0.326 \columnwidth]{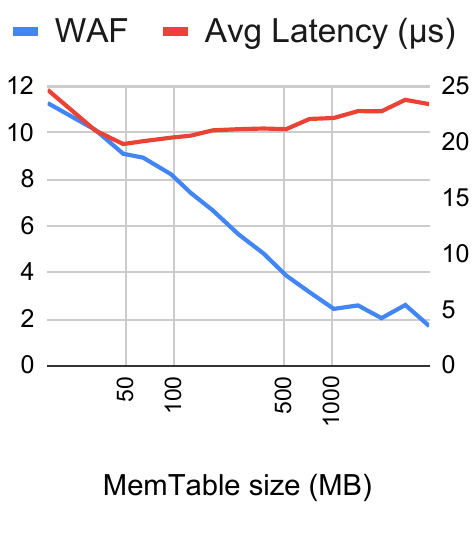}\label{figure:rocksdb:wm}
	}}
	\subfloat[\centering TurtleKV]{{  
	  \hspace{-0.2cm}\includegraphics[width=0.36 \columnwidth]{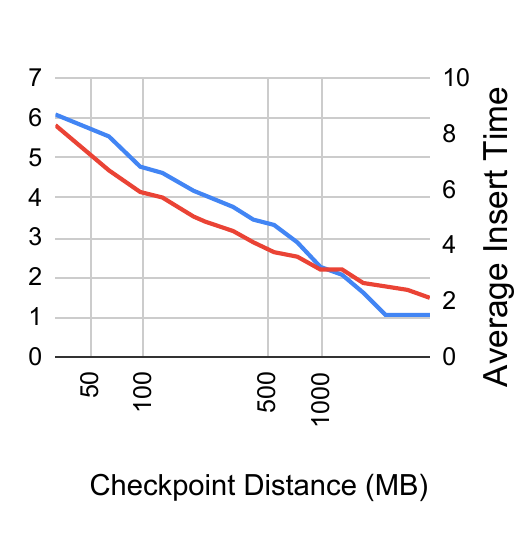}\label{figure:turtlekv:wm}
	}}	
	\vspace{-0.2cm}
   \caption{Write-Memory Trade-off Scaling (N=100M$\times$128B)}
   \vspace{-0.5cm}
\end{figure}

\begin{figure}
  \centering 
	\subfloat[\centering SplinterDB]{{  
	  \includegraphics[width=0.44 \columnwidth]{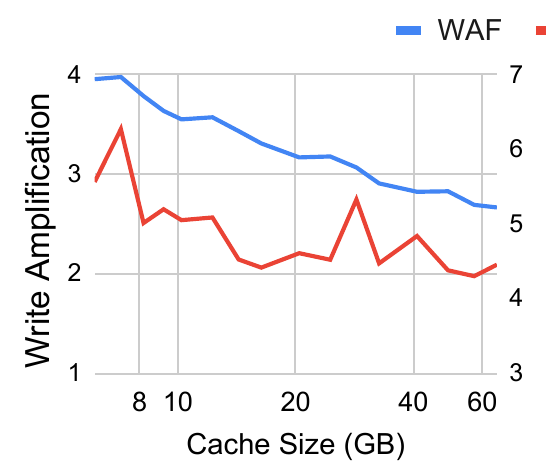}\label{figure:splinterdb:wmcache}
	}}
	\hspace{-0.17cm}\subfloat[\centering TurtleKV]{{  
	  \includegraphics[width=0.44 \columnwidth]{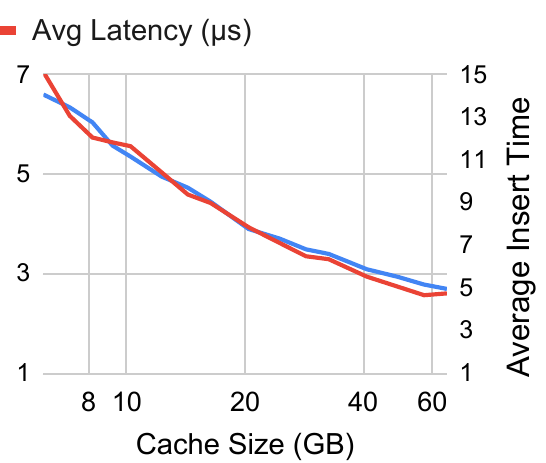}\label{figure:turtlekv:wmcache}
	}}	
	\vspace{-0.2cm}
   \caption{Write/Cache Size Trade-off Scaling (N=500M$\times$128B)
   \textit{Note: SplinterDB sacrifices read performance for faster ingest/updates (see Figure \ref{figure:ycsb:vs_world} \& Figure \ref{figure:ycsb:latency}, $\S$\ref{section:eval:results})}}
   \label{figure:wmcache}
   \vspace{-0.4cm}
\end{figure}

Figure \ref{figure:rocksdb:wm} shows the result of repeatedly running
a uniform random insertion workload against an empty RocksDB database,
while increasing \MemTableSize{} from 16MB up to 4GB.  When 16MB $\le
\MemTableSizeM \le$ 1GB, we see the predicted logarithmic reduction in
write amplification (x-axis is log scale).  However, for
MemTable/SSTable sizes over 1GB, the write optimization effect
flattens out, since the inserted/flushed data never escapes $L_0$
before the workload completes\footnote{Key ranges of SSTable files in
$L_0$ are allowed to overlap, so as not to stall MemTable flushes; at
least 4 files must be present in $L_0$ to trigger a compaction to
$L_1$.}.  Unfortunately this reduction in write volume does not
translate into lower average insert latencies beyond RocksDB's default
value of \MemTableSize{} =64MB.  This underscores the necessity of
optimizing main-memory costs (CPU, memory bandwidth, and internal
synchronization bottlenecks) if one targets I/O reduction as a tuning
mechanism.

\vspace{-0.15cm}
\subsubsection{SplinterDB}  

SplinterDB largely adopts the strategy of selecting a single tuning
point which attempts to offer excellent read and write efficiency by
structuring per-node buffer data pages as a single-level, size-tiered
LSM-tree-like structure, where each sorted run (or \textit {branch})
is stored as a statically built B$^+$-tree.  Quotient Maplets, a
variant of Quotient Filters \cite{pandeyVectorQuotientFilters2021a},
are used in this design to route point queries to the branches/tiers
which may contain a given key.  SplinterDB's STB$^\epsilon$-trees have
configurable branching factor, which is tied to the number of tiers
per level $T$.  Quotient filter bit rate may also be tuned, and thus
STB$^\epsilon$-trees theoretically support a degree of write-memory
tuning via filter arbitrage.

When we attempt to measure the effectiveness of these tuning
parameters in practice, we observe minimal effects from varying
fanout, filter size, and memtable size in SplinterDB.  In fact the
only parameter we observe to have a significant effect on end-to-end
performance and I/O amplification is cache size, as shown in Figure
\ref{figure:splinterdb:wmcache}.  We increased the data set size to
500M$\times$128B for this test to show more of the tuning range, since
at N=100M WAF was nearly optimal ($\approx{1}$), even at small cache
size. SplinterDB's size-tiered lazy compaction policy achieves very
low write amplification at small cache sizes, and for the most part
average insert latency tracks with WAF.  However, we find SplinterDB's
dynamic range of WM tuning to be limited, RM and WM cannot be
independently targeted, and the task-parallel ($\S$\ref{section:design:parallelstrategy}) compaction strategy
tends to generate a high variance in insert latencies.

\vspace{-0.15cm}
\subsubsection{TurtleKV}

For comparison, we show \TurtleKV{}'s results for both N=100M
write-buffer size scaling and N=500M cache size-scaling workloads in
Figures \ref{figure:turtlekv:wm} and \ref{figure:turtlekv:wmcache}
respectively.  These results show that \TurtleKV{}'s primary data
structure, the \TurtleTree{}, efficiently reduces write cost at a rate
of $-O(\log(M))$, where $M$ is the memory spent on write-optimization,
like level-tiered \LSMtree{}-based RocksDB.  Furthermore, because of the
system design ideas elaborated in the following sections, \TurtleKV{}
is able to turn WAF reduction into higher performance over a very wide
tuning parameter range, similar to WiredTiger. When configured for
write optimization, \TurtleKV{} keeps pace with SOTA engine
SplinterDB, and because \TurtleKV{}'s data-parallel strategy for
in-memory compactions is able to achieve even load balancing across
processors, its scaling curve is much smoother than SplinterDB's
(Figure \ref{figure:wmcache}).

\vspace{-0.15cm}
\subsection{Towards Dynamic Tunability: Key Ideas}\label{keyIdeasSection}

Table \ref{table:db_comparison} expresses the three high-level requirements
which drive the design of \TurtleKV{}.  In this section we provide a
brief sketch of the key ideas which address each requirement.  These ideas
will be elaborated in the following sections.

\begin{figure*}
  \centering
  \subfloat[\centering Logical Structure]{{  
    \includegraphics[width=0.88 \columnwidth]{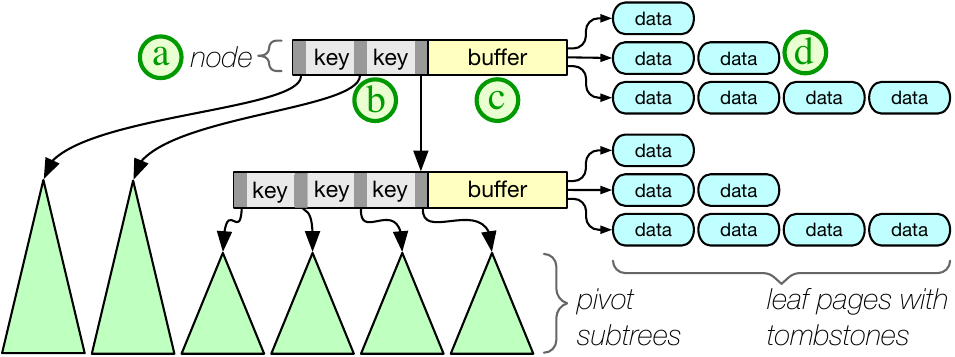}\hspace{2cm}
    \label{turtleTreeFigure}
  }}
  \subfloat[\centering Page Layout]{{  
     \includegraphics[width=0.88 \columnwidth]{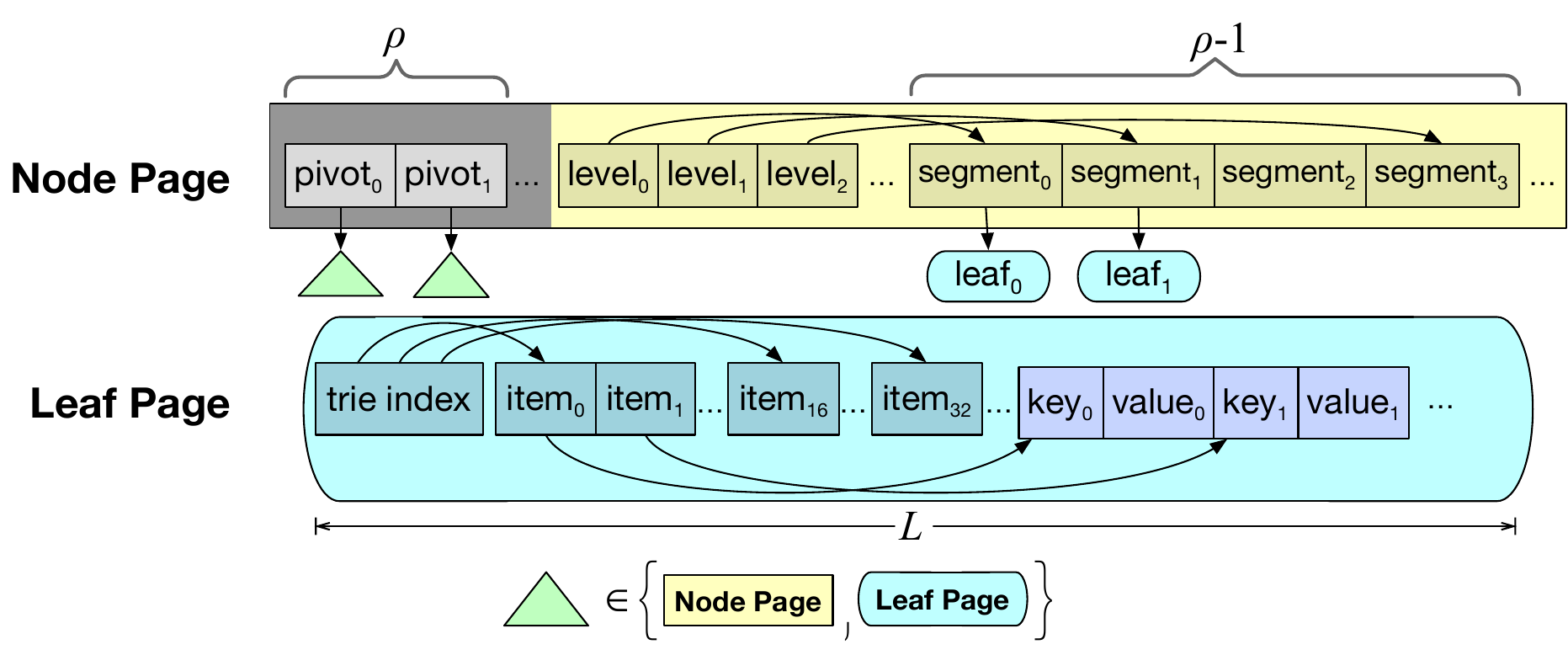}
		\label{turtleTreeDetailFigure}
    	      }}
	\vspace{-0.25cm}
	
	 \caption{The \TurtleTree{} data structure.}
	\vspace{-0.4cm}
\end{figure*}

\vspace{-0.15cm}
\subsubsection{Data Structure With Efficient WM Trade-off}

We start with the \STBepsilontree{}, observing that its desirable
properties derive from the trunk/branch separation, whereas the
properties we would like to change derive from their compaction and
flush policies.  We coin the term \Bepsilonplustree{} to describe the
class of dictionaries which retain the trunk/branch separation, but
which leave open the questions of flush/compaction policy and how (if
at all) branches are structured within their referent trunk node(s)
(Figure \ref{bEpsilonPlusTransformation}).  Our first key idea is to
instantiate the \Bepsilonplustree{} with a balanced,
\textit{level-tiered} compaction policy which organizes branches into
sub-levels within a trunk node, as shown in Figure
\ref{turtleTreeFigure}.  This structure we name as the
\TurtleTree{}.

\vspace{-0.15cm}
\subsubsection{System Design With No In-Memory Bottlenecks}

\TurtleKV{} adopts a design similar to \LSMtree{}-based systems like
RocksDB, with a write-ahead log (WAL), in-memory component (MemTable),
and log-structured on-disk component (Checkpoint, a \TurtleTree{}).
The primary bottleneck in this design comes from the $\log(N)$
amplification when merging MemTables into the current Checkpoint.  We
observe that the most CPU-intensive part of this process is the key
comparisons for level-to-level merging, and the most
memory-bandwidth-intensive part is the copying of key-value data to
serialize compacted data at its new location.  Our second key idea is
to use data-parallel algorithms for level merging, and fork/join style
multi-threaded leaf page serialization for Checkpoint writes.

\vspace{-0.15cm}
\subsubsection{First-Class Write-Memory (WM) Tuning Knob}

Inspired by our RocksDB case study, we identify MemTable size
\MemTableSize{} as our WM tuning knob.  This defines the distance, in
total bytes of updates, between Checkpoints.  To ensure that
increasing checkpoint distance does not degrade in-memory query
performance, we build a single large MemTable between checkpoint
updates.  To keep tail latencies under control, and to avoid affecting
the on-disk structure of checkpoints, we break finalized MemTables
into leaf-page sized batches by doing a key-order scan, applying the
batches to the Checkpoint \TurtleTree{} one at a time in-cache until
the MemTable is drained.  This is our third key idea, which we call
checkpoint distance tuning with Big MemTables.

%% file: figures/fig_chivslazytuning.tex
\begin{figure}[b]
	\centering
    \vspace{-0.5cm}
	 \includegraphics[width=0.48 \textwidth]{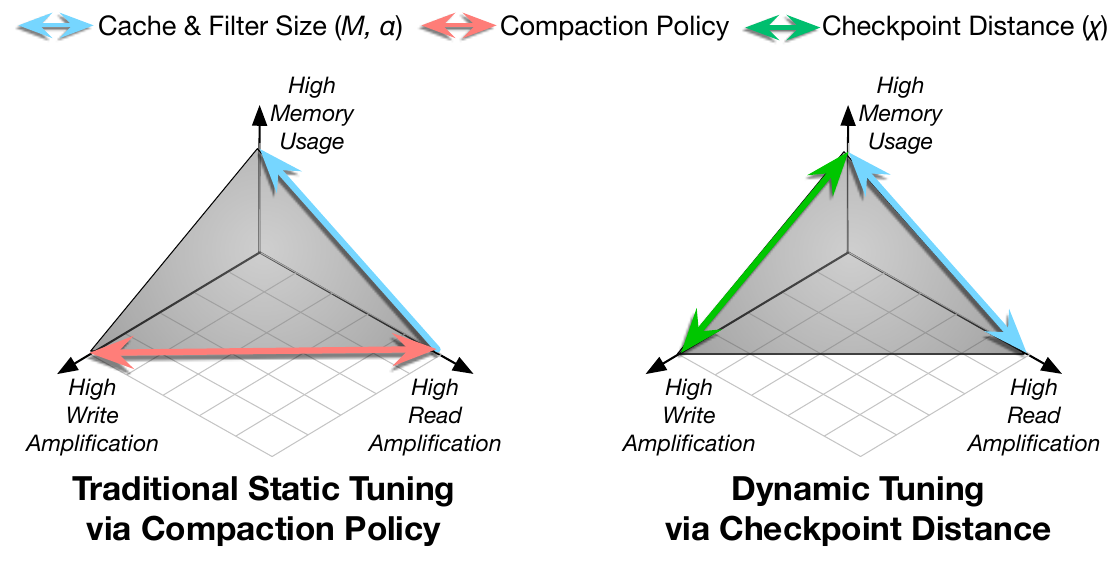}
     \caption{RUM as a three-dimensional resource space}
      \label{chiVsLazyTuning}      
    %\vspace{-0.5cm}
\end{figure}

%% file: figures/table_db_comparison.tex
\begin{table}[b]
  \centering
  \caption{The three key requirements for dynamic write-memory trade-off tuning in key-value storage engine design.}
    \label{table:db_comparison}
\resizebox{\columnwidth}{!}{
  \begin{tabular}{l c c c c}
    \toprule
     \hphantom{}
    & \hphantom{} \large{\textbf{WiredTiger}} \hphantom{}
    & \hphantom{} \large{\textbf{RocksDB}} \hphantom{}
    & \hphantom{} \large{\textbf{SplinterDB}} \hphantom{}
    & \hphantom{} \large{\textbf{TurtleKV}} \hphantom{}
    \\
    \midrule
    \begin{tabular}{@{}c@{}}
	    \large{\textit{Data Structure supports}} \\ 
        \large{\textit{efficient WM trade-off?}}
    \end{tabular}    
    \vspace{0.2cm}
    & \LARGE{\color{red}\xmark}
    & \LARGE{\cmark}
    & \LARGE{\cmark}
    & \LARGE{\cmark}
    \\
    %\midrule
    \begin{tabular}{@{}c@{}}
	    \large{\textit{System Design has no}} \\ 
        \large{\textit{in-memory bottlenecks?}}
    \end{tabular}    
    \vspace{0.2cm}
    \textit{}
    & \LARGE{\cmark}
    & \LARGE{\color{red}\xmark}
    & \LARGE{\cmark}
    & \LARGE{\cmark}
    \\
    %& & & & & \\
    %\midrule
    \begin{tabular}{@{}c@{}}
	    \large{\textit{First-class write-memory}} \\ 
        \large{\textit{trade-off tuning knob?}}
    \end{tabular}    
    \textit{}
    & \LARGE{\cmark}
    & \LARGE{\cmark}
    & \LARGE{\color{red}\xmark}
    & \LARGE{\cmark}
    \\
    \bottomrule
  \end{tabular}
  }
  %\vspace{0.2cm}
\end{table}

%% file: sections/3-turtle-trees.tex
\section{\TurtleTree{}s}
\label{sec:design}

In this section we describe the \TurtleTree{} data structure, its core
algorithms (batch update, get, and scan), and how it leverages
checkpoint distance as a WM trade-off tuning mechanism.

\subsection{\TurtleTree{} Structure}
\label{sec:design:turtles}

\TurtleTree{}s are B$^{\epsilon +}$-trees ($\S$\ref{keyIdeasSection})
whose per-node update buffer \gcircled{c} is organized into a fixed number of
levels \gcircled{d}, as in Figure \ref{turtleTreeFigure}.  Figure
\ref{turtleTreeDetailFigure} shows the layout of
\TurtleTree{} node and leaf pages.

\vspace{-0.15cm}
\subsubsection{Leaf Pages}
Leaf pages contain a sorted list of key-value pairs.  To achieve high
I/O utilization during updates and scans, leaf pages are sized several
orders of magnitude larger than interior nodes.\footnote{The
configuration we use is 4KB node pages and 32MB leaf pages.} In order
to guarantee low latencies during point queries, leaf pages are
internally structured to allow small slices, or \textit{sharded
  views}, to be cached and re-used to find queried keys. Leaves also
include a cache-oblivious, byte-addressed binary trie index at the
front of the page.  This index contains every $k^{th}$ key from the
leaf, $k\in\{1..32\}$.  The trie is made cache-oblivious by laying out
nodes in Van Emde Boas order
\cite{benderCacheobliviousStreamingBtrees2007a} to maximize locality
of reference along any given query path, minimizing CPU cache misses
for in-memory queries.

\subsubsection{Node Pages}\label{section:turtletree:nodepage}
Each (interior) node \gcircled{a} is split into roughly equal
halves,\footnote{The intra-page regions in Figure
\ref{turtleTreeDetailFigure} are not shown to scale.} with one half
devoted to pivots (dark gray \gcircled{b}) and the other to the update
buffer (yellow \gcircled{c}).  Within a node page, the update buffer
region contains pointers to \textit{segments}, which are stored as
leaf pages with the addition of tombstone records for deleted
keys. The segments in a buffer are stratified into levels of
exponentially increasing size.  Keys are required to be
\textit{in-order} and \textit{unique} within a level.  The first level
contains at most a single segment, the second level at most two, the
third level four, then eight, etc.  All levels start out
\textit{vacant} and can be in either a \textit{vacant} or
\textit{occupied} state. For a given number of pivots $\rho$ in a
node, the number of levels is limited to $\lceil\log_{2}{\rho}\rceil$,
and the total number of segments in all levels of the buffer limited
to at most $\rho - 1$.

Node buffers maintain certain metadata to speed up various operations.
For each segment, we store a bit set \textit{activePivots} of the
pivots addressed by at least one active key in the
segment,\footnote{We say that a pivot is \textit{addressed by} a key
if the pivot's key range contains that key.} and a compressed sparse
array \textit{flushedPivots} of key index upper bound values for each
pivot to which keys from the segment have been flushed.  When keys are
removed from a level as part of a flush operation, the segment pages
are not modified; instead, the flushed upper bound array is updated
for that pivot so it points to the segment's first non-flushed key in
the flushed pivot key range.  When all keys addressed to a given pivot
have been flushed from a segment, the segment's active bit is cleared
for that pivot, and the flushed upper bound is reset to zero.  If this
causes the active set to become empty, the segment is removed from the
node, possibly allowing its storage to be
garbage-collected.\footnote{Operations such as node splits during a
batch update or taking a snapshot of an existing \TurtleTree{} may
introduce additional references to a given page.}

Node pages also maintain a count of the number of bytes of buffered
updates in the key range of each pivot.  This is used to decide when
and how to perform flush operations.  The default policy is to flush
one leaf-sized batch after each batch insert, to the pivot with the
highest \textit{pendingBytes} count, if possible.

\subsection{\TurtleTree{} Operations}
\label{sec:design:ops}

\subsubsection{Batch Updates}\label{section:turtle_tree:ops:update}
\TurtleTree{}s are designed to support efficient batched updates;
individual key modifications can be logged/buffered outside the tree
and applied as a batch once enough have
accumulated. \textit{\TurtleTree{}s do not directly implement
  single-key \textit{Put} or \textit{Delete}}.

%\textbf{Batch Updates.} Turtle Trees are designed to support efficient
%batched updates; individual key modifications can be logged/buffered
%outside the tree and applied as a batch once enough have accumulated.
%As such, Turtle Trees do not directly implement single-key
%\textit{Put} or \textit{Delete}.

When the tree is empty, or consists of a single leaf, most operations
are trivial.  Batch update on a leaf consists of doing a merge/compact
on the new and old data; if the result is small enough to fit in a
leaf page, then the tree remains a single leaf.  If not, then a new
node is created to parent the two leaves.  The root node is the only
node in the tree allowed to have an arbitrarily small number of child
pivots, so long as it is at least 2.

\begin{figure}[t]
	\centering
	\includegraphics[width=0.9 \columnwidth]{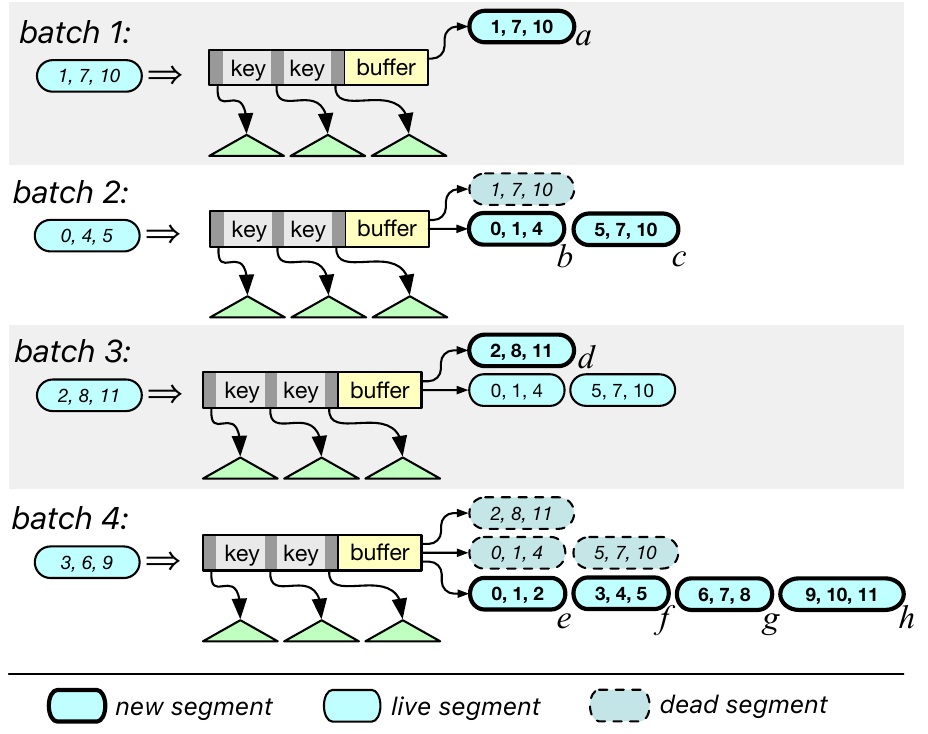}
	\vspace{-0.3cm}
	\caption{Insertion of Batch Updates in \TurtleTree{}s.}
    \label{turtleTreeBatchUpdateFigure}
	\vspace{-0.5cm}
\end{figure}

To perform a batch update on a node, we first insert the incoming
batch into the node's buffer.  To illustrate how this works, we work
through a simple example, shown in Figure
\ref{turtleTreeBatchUpdateFigure}.  In this example, we process four
batches, containing keys $\langle 1,7,10 \rangle$, $\langle 0, 4, 5
\rangle$, $\langle 2, 8, 11 \rangle$, and $\langle 3, 6, 9 \rangle$
respectively.  The node's buffer is initially in an \textit{empty}
state, i.e. all levels are vacant.

When the first batch $\langle 1,7,10 \rangle$ is inserted, we look at
the first level of the buffer to see if it is vacant or occupied.  It
is vacant, so we can simply insert the batch as a new segment page
\textit{a}.  At the second batch, $\langle 0, 4, 5 \rangle$, we again
check the first level, but this time it is occupied.  Thus, we must
perform a merge of the keys in segment \textit{a} and \textit{batch 2}
to produce the sequence $\langle 0, 1, 4, 5, 7, 10 \rangle$, which is
split into two segments \textit{b} $\langle 0, 1, 4 \rangle$ and
\textit{c} $\langle 5, 7, 10 \rangle$.  The third batch $\langle 2, 8,
11 \rangle$ becomes a new segment \textit{d} occupying the first
level, just as the first batch.  When the fourth batch $\langle 3, 6,
9 \rangle$ is inserted, it must be merged with both levels one and
two, since both are occupied.  This cascade of merges produces the
final sorted key sequence $\langle 0, 1, 2, 3, 4, 5, 6, 7, 8, 9, 10,
11\rangle$, resulting in a new level comprised of segments \textit{e}
$\langle 0, 1, 2 \rangle$, \textit{f} $\langle 3, 4, 5 \rangle$,
\textit{g} $\langle 6, 7, 8 \rangle$, and \textit{h} $\langle 9, 10,
11 \rangle$.

After inserting a new batch into the buffer, if there are at least $L$
(the size of a leaf page) buffered bytes of updates addressed to one
of the pivots, then we perform a flush.  This involves picking the
pivot with the most buffered data, and removing $L$ bytes of updates
from the buffer in the target key range.  Extracting a batch to flush
may involve an in-memory merge/compaction, as the data may be spread
across multiple levels.  In the prior example, if we perform a flush
after \textit{batch 3}, supposing the target subtree's key range to be
$2..6$, we would need to merge data from all three segments
(\textit{b, c, d}) to produce the batch $\langle 2, 4, 5\rangle$.  As
mentioned in $\S$\ref{sec:design:turtles}, durable data in update
buffer segments is ``removed'' by modifying the in-node active key range
metadata ($\S$\ref{section:turtletree:nodepage}).  This reduces write
amplification by avoiding the need to update the segment page(s).

\input{figures/table_cost_analysis}

After a batch has been recursively flushed to a sub-tree, a split or
join of the corresponding pivot may be required.  All non-root nodes
must be kept above a configured minimum size to remain viable.  In addition, we
maintain an invariant that the total amount of buffered data per node
never exceeds $L \cdot (\rho-1).$ If the buffered data does ever
exceed this amount, it will be because a new batch of size $L$ was
just inserted, resulting in there being at least $L \cdot \rho$ total
bytes.  In this case, we will always have enough to flush a batch to
some sub-tree to restore the invariant. Node joins are simpler than
splits, since we can assume the invariant to hold for each node being
joined.  Since the key-range of two sibling nodes is guaranteed not to
overlap, the invariant is preserved by simply concatenating the
contents of nodes to join them.  The same is not true, in general, of
splits.  If after splitting a node as evenly as possible, we find that
the update buffer size invariant is violated for one of the new
smaller nodes, we perform another flush on that node to restore it.
Note that we can be sure that if this case arises, it will only be for
one side, not both; otherwise, the invariant must have been false
before the split.  An AMQ filter per leaf page is built and flushed
when batch updates are finally written to storage.

\subsubsection{Queries}\label{section:turtle_tree:ops:query}
Point and range queries are the same for \TurtleTree{}s as for
B$^\epsilon$-trees, with the addition of chasing pointers to segment
page(s) in the buffer (instead of just reading its contents).  We use
the \textit{activePivots} bit sets stored for each segment to narrow the
search to only those which might contain the key.  Further pruning of
the set of candidate pages to search is done by querying a Bloom or Quotient
filter \cite{pandeyVectorQuotientFilters2021a} associated with each
segment/leaf ($\S$\ref{section:turtle_kv:basic:query_path}).

After finding a key in a segment, we compare its position in the leaf
to the flushed upper bound for its corresponding pivot; if this
position is less than the upper bound (i.e., it has been marked as
flushed), we act as if the key is not present and continue the search.

\textit{Delete}s are performed alongside \textit{Put} operations as
part of a batch update, by setting a key's value to the special
\texttt{DELETED} tombstone value.  This value is preserved in leaves
until the bottom of the tree is reached; at this point, the key is
removed from the merged output.

\subsection{\TurtleTree{} Cost Analysis}
\label{sec:design:checkpoint}

\subsubsection{Update I/O cost}
Individual key updates travel from root to leaf of the tree as batch
updates and flushes take place.  Since there are $O(\log_{2}{\rho})$
levels within the buffer of each node, and the tree height is
$O(\log_{\rho}{(N/L)})$, each key update is written
$O(\log_{2}{(N/L)})$ $ = O( \log_{2}{\rho} \cdot \log_{\rho}{( N / L
  )})$ times, regardless of page size or branching factor
$\rho$.  \footnote{Note that in order to correctly analyze the cost of
a key update, we need only consider its initial journey from
root-to-leaf, since each time a key's page is re-written after it
reaches the bottom of the tree, we can amortize that cost against the
incoming batch, which represents an equal number/size of keys ending
their root-to-leaf journey.}  If we define the transfer size $B$ in
units equal to a single key update, the I/O cost of a key update is
$O(B^{-1} \cdot \log_{2}{(N/L)})$.

\vspace{-0.15cm}
\subsubsection{Query cost}
Point queries (\textit{Get}) differ in their I/O cost depending on
whether or not the key is found.  Here we give a worst-case bound for
the general case; in practice this can be dramatically improved
through the addition of per-leaf filters, as described in
$\S$\ref{section:turtle_kv:basic:query_path}.  The cost analysis here
also depends on the specific I/O model used; if we are counting total
page loads, as in the Disk Access Model (DAM), then a single-key
(point) query must load all node pages and buffer segments along a
single root-to-leaf path: $O( \log_{\rho}{(N/L)} + \frac{L}{B}\cdot
\log_{2}{(N/L)} )$ or just $O( \frac{L}{B}\cdot \log_{2}{\frac{N}{L}}
)$.

Like other B-tree variants, range queries of length $k$ on Turtle Trees cost a point
query (to find the starting key) plus $O(k/B)$
IOPs.  There is also an in-memory, per-key cost of
$\log_2{(\log_2{(N/L)})}$ to merge sorted runs, which is the same as a
level-tiered LSM-tree.  This per-key cost expression is derived via
the use of a binary min-heap to select which of the $\log_2{(N/L)}$
levels of the tree contains the least-ordered key.

\vspace{-0.15cm}
\subsubsection{Write-Memory Trade-off Efficiency}
\begin{figure}[t]
	\centering \includegraphics[width=0.85
      \columnwidth]{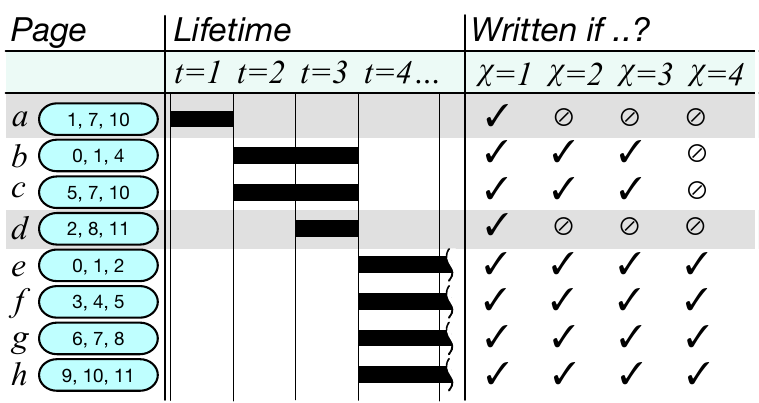}
	\vspace{-0.15cm}
	\caption{The effect of checkpoint distance (\checkpointdistance{}) on write
          amplification.  (Pages shown are from the example
          in Figure
          \ref{turtleTreeBatchUpdateFigure}.)}\label{turtleTreeChiEffectFigure}
	\vspace{-0.45cm}
\end{figure}

The WM tuning knob for \TurtleTree{}s is a parameter \textit{chi}
($\chi$), for \textit{checkpoint distance}, which determines the
number of batches applied to a \TurtleTree{} in-memory before flushing
changed pages to storage as a new checkpoint.  At $\chi=1$, every time
we perform a batch update on the tree, all modified pages must be
written back out.  At $\chi=2$, we defer the writing of new/modified
pages until two batches have been applied, etc.  It may not be
immediately obvious to the reader that $\chi$ should have any effect
on the write cost of key updates.  To see that it does, let us revisit
the batch insertion example illustrated by Figure
\ref{turtleTreeBatchUpdateFigure}, considering the span of time (as
measured by batch number) that each page is live, shown in Figure
\ref{turtleTreeChiEffectFigure}.  The pages which appear in durable
storage at a given time step are those whose lifespan overlaps a given
column in the middle section (\textit{Lifetime}) of the diagram.

Observe that when $\chi=2$, the state of the tree is only externalized
at time steps \textit{2} and \textit{4}, so that the segments
(\textit{a}, \textit{d}), which are live only for steps 1 and 3, are
never written.  Also observe that when $\chi=2$, keys 1, 7, and 10
first appear in stored pages at level two in segments \textit{b} and
\textit{c}; i.e., they appear to an external observer (who can only
look at the \textit{externalized} or durable states of the tree) to
have \textit{skipped} the first level altogether.  If one considers
\textit{batch 3}, the effect is even more pronounced: the first place
keys 2, 8, and 11 appear when $\chi=2$ is at the third level, in
segments \textit{e}, \textit{g}, and \textit{h}.  If $\chi$ is doubled
again to 4, then all keys in the example appear to skip the first two
levels.

We claim that, in general, the effect of \checkpointdistance{} is that
keys skip the first $\log_2{\checkpointdistanceM}$ buffer levels of
the tree, which reduces the per-key write cost to: $$
O\left(\frac{1}{B} \cdot \left( \log_{2}{\frac{N}{L}} -
\log_2{\chi}\right) \right) = O\left(\frac{1}{B} \cdot
\log_{2}{\frac{N}{\chi L}} \right)$$

The key limitation of \TurtleTree{}'s \checkpointdistance{} tuning
knob, and of RM+WM tuning schemes in general, is that unlike I/O
operations, memory is generally a bounded quantity.  This means higher
\checkpointdistance{} value will spend memory that could have been
used for page and filter caching.  Thus the zero-sum of fixed-size
memory resources is effectively turned into a read-vs-write (RW)
trade-off.  However, unlike RW trade-offs that affect the structure of
stored data, our scheme is flexible and forgiving, since it can be
changed at runtime.

%% file: figures/table_cost_analysis.tex
\begin{table*}[b]
  \centering
  \caption{Per-key, asymptotic costs (Big-$O$) for external memory
    dictionaries. \iffalse{} N = number of inserts, B = I/O transfer (block)
    size, F = level growth factor, T = tiers per level, $0 \le
    \epsilon \le 1$ controls the fraction of a node allocated for
    pivots ($B^\epsilon$) vs buffer ($B^{1-\epsilon}$).  \fi{} Point-Query
    (\textit{Get}) costs are given both for Disk Access Model (DAM)
    and Parallel Disk Access Model (PDAM). \iffalse{} We omit the $L$ (leaf
    size) parameter and assume $\rho = O(B)$, to simplify the
    expressions and to ease comparison across data structures. \fi{} Scans (length $k$) cost a Get plus $\frac{k}{B}$ I/Os for all cases and models.}
    \label{table:cost_benefit}
  \begin{tabular}{l c c c c c c}
    \toprule
    \textbf{Operation} \hphantom{}
    & \hphantom{} \large{\textbf{\TurtleTree{}}} \hphantom{}
    & \hphantom{} \textbf{B$^+$-Tree} \hphantom{xx}
    & \hphantom{} \textbf{LSM (level-tiered)} \hphantom{}
    & \hphantom{} \textbf{LSM (size-tiered)} \hphantom{}
    & \hphantom{} \textbf{B$^\epsilon$-Tree} \hphantom{}
    & \hphantom{} \textbf{STB$^\epsilon$-Tree} \hphantom{}
    \\
    \midrule
    \textit{Put}
    & \boldmath$\frac{1}{B} \cdot \log_2{\frac{N}{\chi}}$
    & $\log_B{N}$
    & $\frac{F}{B} \cdot \log_F{N}$
    & $\frac{1}{B} \cdot \log_T{N}$
    & $\frac{1}{B^{1-\epsilon}} \cdot \log_{B^\epsilon}{N}$
    & $\frac{1}{B} \cdot \log_F{N}$
    \vspace{0.1cm}
    \\
    %& & & & & \\
    \textit{Get} (DAM)
    & \large{\boldmath$\log_{2}{N}$}
    & $\log_B{N}$
    & $\log_{F}{N} \cdot \log_B{N}$
    & $T \cdot \log_{T}{N} \cdot \log_B{N}$
    & $\frac{1}{\epsilon} \cdot \log_{B^\epsilon}{N}$
    & $F \cdot \log_B{N}$
    \vspace{0.1cm}
    \\
    %& & & & & \\
    \textit{Get} (PDAM)
    & \large{\boldmath$\log_{B}{N}$}
    & $\log_B{N}$
    & $\log_B{N}$
    & $T \cdot \log_B{N}$
    & $\frac{1}{\epsilon} \cdot \log_{B^\epsilon}{N}$
    & $F \cdot \log_B{N}$
    \vspace{0.1cm}
    \\
    \bottomrule
  \end{tabular}
\end{table*}

%% file: sections/4-turtle-kv.tex
%\section{TurtleKV: A Dynamic Hybrid Optimized Key-Value Storage Engine}
\vspace{0.1cm}
\section{\TurtleKV{} System Design}
\label{sec:turtle_kv}

In this section we first give an overview of the system architecture
and update/query paths in \TurtleKV{}, then dive deeper into the
design of two of the key ideas it implements: checkpoint distance
tuning with Big MemTables, and parallel (multi-core) acceleration
for \TurtleTree{} (checkpoint) updates.

\vspace{-0.15cm}
\subsection{Basic Operation}

\TurtleKV{} is based on a conventional \LSMtree{} system architecture
employing three high-level components: a write-ahead log (WAL) to
ensure durability of incoming updates, an in-memory index and buffer
of recent data (MemTable), and an on-disk checkpoint for long-term
storage (\TurtleTree{}).  It is implemented in ~32k lines of C++ code,
and is available as open-sourced
software \cite{HttpsGithubcomMathworks2025}.
 
\vspace{-0.15cm}
\subsubsection{Update Path}
Incoming key-value data are copied to an internal memory buffer which
is continuously flushed in the background to the WAL.  Each update is
assigned a unique sequence number so that the database state can be
consistently reconstructed during recovery.  After the key and value
bytes are copied, they are inserted into an in-memory index
(the \textit{active} MemTable) on the user thread performing the
update; as soon as this completes, the update is visible to all
threads in the process.

Each MemTable is limited to a maximum size \MemTableSize{}.  When a
thread attempts to perform a MemTable insertion, but fails because
this limit would be exceeded, the active MemTable
is \textit{finalized}, or made read-only, and a new active MemTable is
created to accept future updates.  The finalized MemTable is handed
off to a background thread, which splits it into 1 or more
leaf-page-sized batches, each of which is applied to the current
checkpoint \TurtleTree{}.  The user thread must wait if the background
thread is already busy processing a previous finalized MemTable.
After the hand-off succeeds, it will also wait until the total number
of finalized MemTables in the system is at most 2.  When the
background thread finishes applying all batches for a MemTable to the
page cache, it hands the list of new pages to a second background
thread (the \textit{writer thread}), which writes the new checkpoint
data to disk.  Once the new \TurtleTree{} pages are durable, the
writer thread swaps the finalized MemTable with a reference to the new
checkpoint which subsumes it, possibly unblocking any application
threads that were waiting for the finalized MemTable count to
decrease.

This multi-stage update pipeline allows high system resource
utilization by allowing the different stages (MemTable
insertion, \TurtleTree{} updates, and checkpoint page flushing) to be
performed at the same time.  By keeping the amount of queued data in
the pipeline to a minimum, it also provides back-pressure so that no
part of the system outpaces the others, in order to avoid unbounded
tail latencies.

\vspace{-0.15cm}
\subsubsection{Query Path}\label{section:turtle_kv:basic:query_path}
Point queries are answered by first consulting the active MemTable.
If the target key is found there, the query terminates and the answer
is returned to the caller.  Otherwise, the $\le$2 finalized MemTables
are consulted in newest-to-oldest order.  If the key is not found
there either, then the most recent checkpoint is queried as described
in $\S$\ref{section:turtle_tree:ops:query}.

\TurtleKV{} supports either Bloom or Quotient filters (selected at
compile-time) to optimize point queries.  Each leaf page, whether at
the bottom of the \TurtleTree{} or referenced from a node buffer, has
an associated filter page.  Whenever the query algorithm demands that
a leaf be searched for a given key, the associated filter is first
queried, and only if it gives a positive response is the leaf
loaded/checked.

When a leaf is queried, we first attempt to pin the whole leaf page in
the cache; this will fail if the page isn't in memory, without
performing any I/O.  If the whole leaf is not cached, then we access
the first ~64KB-sized slice of the page, which contains the header and
trie index.  From this information, we are able to determine which 4KB
slice will contain the target key, if present.

\subsection{Parallelization of \TurtleTree{} Updates}

\subsubsection{The necessity of parallelized updates}
\TurtleTree{} checkpoint update is the most expensive operation in the
\TurtleKV{} update path by a substantial margin.  Therefore it is the
primary target of our optimization efforts to avoid in-memory
bottlenecks, which is one of the 3 high-level requirements
for \TurtleKV{} ($\S$\ref{section:background_key_ideas},
Table \ref{table:db_comparison}).

To put this in context, while running YCSB workloads on 16 concurrent
threads, we measured the average MemTable insertion time
in \TurtleKV{} to be approximately 1.5 $\mu{}s$, for a combined rate
across all threads of slightly over 10 million inserts per second.  In
our end-to-end testing, we find that this outpaces the overall load
throughput of the fastest databases (SplinterDB and TurtleKV) by
2-5$\times$.  For the ``small'' records in our tests (8 byte keys + 120
byte values = 128 bytes), this equates to roughly 1.34 GB/s of load on
the WAL, around half the bandwidth of the test SSD.  For ``large''
records (8 byte keys + 992 byte values = 1000 bytes), MemTable
insertion only becomes the bottleneck if the WAL can exceed ~9.4 GB/s.
Since \TurtleKV{}'s WAL flushing implementation is easily able to
match or exceed the MemTable insertion rate, we are left with
checkpoint update as the potential bottleneck in the
system.\footnote{Details of the WAL are omitted here for brevity; for
more details, see \texttt{https://github.com/mathworks/turtle\_kv}.}

The challenge of \TurtleTree{} updates is that they represent a
potentially logarithmic amplification of work as compared to earlier
pipeline stages, due to the $\log(N/L)$ levels through which update
records must be merged/compacted to reach the leaf level.  There is
simply no way to keep up with load generated by earlier stages of the
update path if we do not employ some kind of in-memory parallelization
strategy.

\vspace{-0.15cm}
\subsubsection{Task-parallel vs Data-parallel}\label{section:design:parallelstrategy}
There are two basic approaches to this: allow concurrent updates to
take place in different locations within the same tree, or somehow
parallelize the work of sequentially applied batch updates.  The
former approach (task-parallel) is used by SplinterDB; \TurtleKV{}
takes the latter approach (data-parallel).  The reason is twofold:
first, there are many complications involved in implementing
concurrent updates to \Bepsilonplustree{}s, and these are a superset
of the complexities one faces when implementing
concurrent \Bplustree{}s, a non-trivial task because of the competing
concerns of root-to-leaf traversal for insertion, and leaf-to-root
split/merge for re-balancing.  Second, adopting a data-parallel 
approach affords better load balancing across processors, 
leading to more consistent update performance with less variability.

\vspace{-0.15cm}
\subsubsection{Details of parallel updates}
We observe that the most CPU-intensive operations in \TurtleTree{}
batch update are the key comparisons required to merge/compact level
segments when a node receives an incoming batch from its parent and
flushes an outgoing batch to a subtree.  Merging multiple ordered
sequences is easily done in parallel using
multi-selection \cite{deoOptimalParallelAlgorithm1994} on each
sequence to determine the input and output ranges for different
processors.  By implementing this algorithm, we are able to scale the
CPU work to multiple processors while ensuring the correctness of our
implementation through simplicity of the code.

We also observe that the most memory-bandwidth expensive operation
during checkpoint update is copying key-value data while serializing
new compacted pages prior to writing them to storage.  Our approach to
optimizing this is reduce the total amount of data that must be copied
by deferring all serialization until right before the checkpoint is
finalized, and to serialize all leaf pages (and their corresponding
filters) in parallel on different threads.

\subsection{Checkpoint Distance \& Big MemTables}

\subsubsection{Preliminaries: MemTable Implementation}
The requirements for the MemTable data structure are scalability to
high numbers of concurrent threads, low space overhead, ability to
perform key-order scans, and CPU cache friendliness.  There are
several possible data structures which meet these criteria; for
example, Wang et al. \cite{wangBuildingBwTreeTakes2018} compare
Bw-Tree, OpenBW-Tree, SkipList, Masstree, \Bplustree, and Adaptive
Radix Tree (ART) implementations for in-memory, multi-threaded
performance and efficiency.  \TurtleKV{} implements ARTs for this
purpose, because they outperform all other data structures in this
comparison, and offer low memory usage and high CPU-cache hit rate
across a variety of workloads.  However, \TurtleKV{} is careful to
abstract this implementation choice so that other data structures can
be swapped in if desired.  For comparison, RocksDB uses concurrent
SkipLists as the default MemTable, WiredTiger uses a combination
of \Bplustree{} and SkipLists for in-memory indexing, and SplinterDB
uses concurrent \Bplustree{}s as its MemTable and branches.
The \TurtleKV{} ART implementation is roughly the same size (lines of
code) as RocksDB's SkipList.

\vspace{-0.2cm}
\subsubsection{Checkpoint Distance: A Design Dilemma} \TurtleKV{}'s
write-memory (WM) tuning knob is checkpoint distance; this is the
total size of the updates between two checkpoints.  This can be varied
at runtime to target different levels of write-optimization at the
expense of higher memory footprint devoted to updates.  Recall from
our case study ($\S$\ref{case_study:rocksdb}), that we find there are
two knobs offered by RocksDB to trade write amplification directly
against memory: MemTable size \MemTableSize{} and MemTable
count \MemTableCount{}.  The theoretical reason behind their
effectiveness is that they force the first $O(\log_F(M))$ compactions
for a given key $x$ to happen in memory, before that key is ever
written to an SSTable.  Varying either parameter while holding the
other constant is equally effective, since $M
= \MemTableSizeM \cdot \MemTableCountM$.  The reason to prefer
increasing \MemTableSize{} to \MemTableCount{} as a tuning knob is
that MemTables' contribution to query cost in a RocksDB-like design is
$O(\MemTableCountM \cdot \log(\MemTableSizeM))$, assuming a MemTable
data structure that supports optimal logarithmic query cost.

If we apply this same logic to \TurtleKV{}'s design, however, we
encounter a dilemma.  We want the on-disk structure of
checkpoint \TurtleTree{}s to be independent of checkpoint distance, so
that we get a true WM tuning knob; this is possible only if we
keep \MemTableSize{} constant and vary the number of
batches \MemTableCount{} in between checkpoints.  Thus we have a
tug-of-war between the concerns of in-memory (MemTable) queries on
newer data, and queries against checkpoints (\TurtleTree{}s) in the
face of varied past tuning point selection.

\vspace{-0.2cm}
\subsubsection{The Solution: Big MemTables}
We resolve this tug-of-war by setting the size limit of a MemTable
when it is created to the current checkpoint distance, and breaking up
finalized MemTables into many leaf-page sized batches in order to
update the \TurtleTree{}.  We do this via a least-to-greatest
key-ordered scan of the MemTable.  This design is called Big MemTables
in \TurtleKV{}.

Big MemTables help to optimize \TurtleKV{}'s performance in multiple
ways.  First, by scaling up the size of \TurtleKV{}'s in-memory index
on recent updates, we speed up queries on ``hot keys.''  Second, by
allowing a larger set of updates to accumulate in memory before doing
a scan to break the MemTable into batches, we increase the skew in key
distribution when applying batches to the checkpoint.  \TurtleTree{}
batch update ($\S$\ref{section:turtle_tree:ops:update}), which flushes
batches down the tree as soon as the minimum size threshold for a
subtree is reached (not when the buffer fills up), transparently
leverages this key skew to move data down the tree faster, so long as
doing so will not hurt efficiency (i.e., so long as we preserve the
property that merges are always performed on equal-sized inputs).  We
note that SplinterDB's flush-then-compact policy achieves a similar
effect, though it has the downside of increased space amplification
due to the fact that a flushed branch may end up referenced in many
different subtrees despite only containing a small amount of live data
for each reference.

Finally, Big MemTables further motivate our selection of ARTs as
MemTable data structure.  Because the structure of radix trees allow
for segments of indexed keys to be shared, it is possible for the
per-key overhead of ARTs to fall below the average key
length \cite{leisAdaptiveRadixTree2013}; this efficiency is often
increased as the total set of keys in an ART scales.  Thus by
increasing \MemTableSize{} and then splitting buffered updates into
key-skewed batches, we not only play to the strengths
of \TurtleTree{}s, but to those of ARTs as well.

%% file: sections/5-evaluation.tex
\section{Evaluation}
\label{section:eval}
In this section, we explore the end-to-end performance of \TurtleKV{},
comparing it to WiredTiger, RocksDB, and SplinterDB.  We provide data
to support the claims we make about \TurtleKV{} and its novel data
structure \TurtleTree{}s, and our claims about the necessity of
write-memory (WM) trade-off tuning more generally.

\begin{figure*}
  \centering
  %\vspace{-0.75cm}
  \subfloat[KOps/Second, 48GB cache]{{
    \includegraphics[width=0.45 \columnwidth]{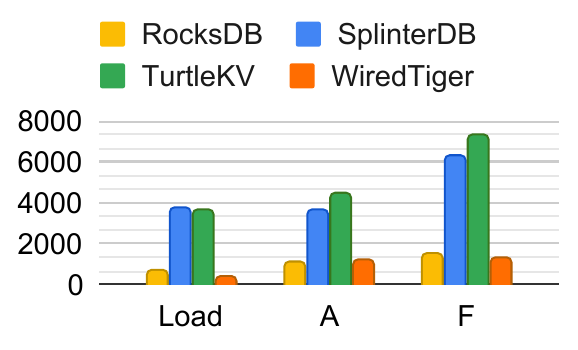}\label{figure:ycsb:vs_world:wr_48gb}
    %\vspace{-0.2cm}
  }}
  \subfloat[KOps/Second, 48GB cache]{{
    \includegraphics[width=0.45 \columnwidth]{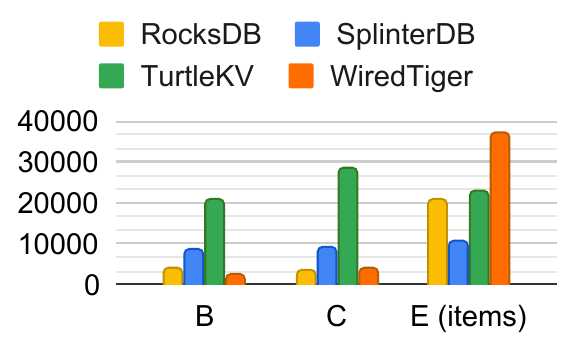}\label{figure:ycsb:vs_world:rd_48gb}
    %\vspace{-0.2cm}
  }}
  \subfloat[KOps/Second, 16GB cache]{{
    \includegraphics[width=0.45 \columnwidth]{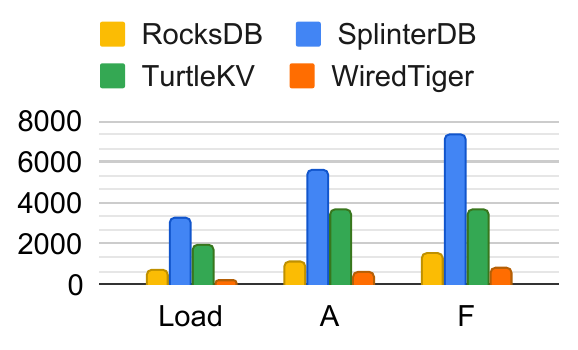}\label{figure:ycsb:vs_world:wr_16gb}
    %\vspace{-0.2cm}
  }}
  \subfloat[KOps/Second, 16GB cache]{{
    \includegraphics[width=0.45 \columnwidth]{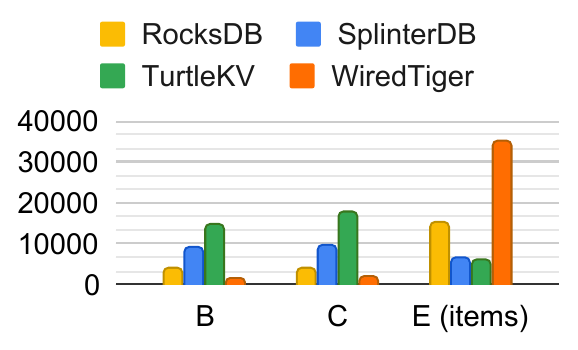}\label{figure:ycsb:vs_world:rd_16gb}
    %\vspace{-0.2cm}
  }}
  %\vspace{-0.25cm}
  \caption{YCSB, N=400M$\times$128b}
  \label{figure:ycsb:vs_world}
  %\vspace{-0.25cm}
\end{figure*}

\begin{figure*}
  \centering
  %\vspace{-0.75cm}
  \subfloat[KOps/Second, 48GB cache]{{
    \includegraphics[width=0.42 \columnwidth]{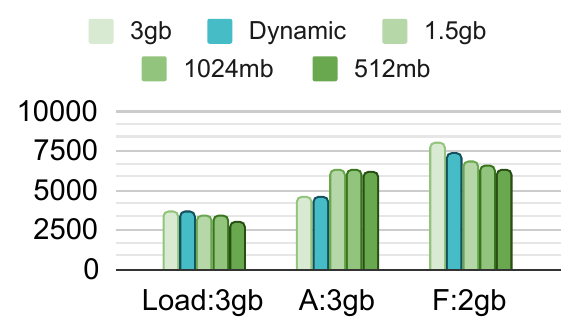}\label{figure:ycsb:vs_turtle:wr_48gb}
  }}
  \subfloat[KOps/Second, 48GB cache]{{
    \includegraphics[width=0.38 \columnwidth]{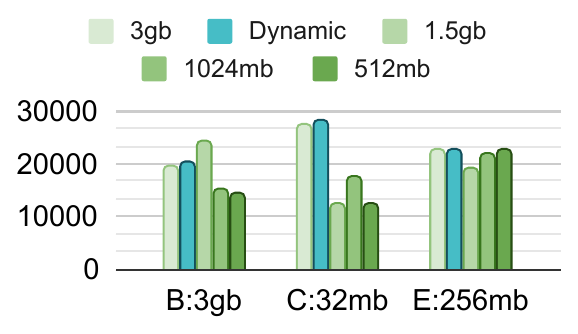}\label{figure:ycsb:vs_turtle:rd_48gb}
  }}
  \subfloat[KOps/Second, 16GB cache]{{
    \includegraphics[width=0.42 \columnwidth]{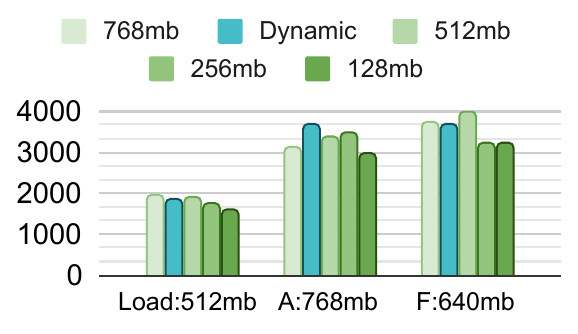}\label{figure:ycsb:vs_turtle:wr_16gb}
  }}
  \subfloat[KOps/Second, 16GB cache]{{
    \includegraphics[width=0.38 \columnwidth]{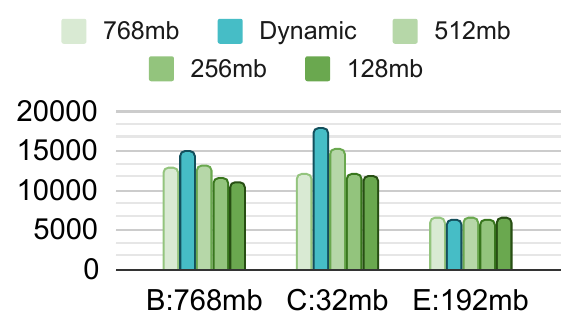}\label{figure:ycsb:vs_turtle:rd_16gb}
  }}
  \subfloat[WAF vs Checkpoint Dist. (MB)]{{
     \includegraphics[width=0.4 \columnwidth]{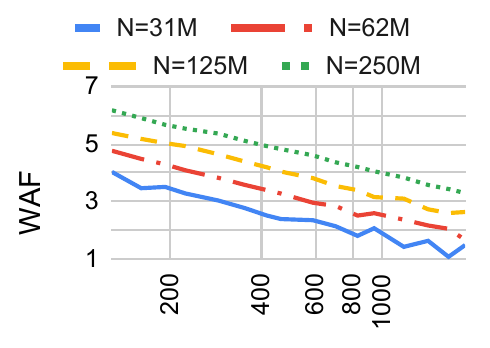}	
     \label{figure:ycsb:vs_turtle:scale_invariant}
  }}
  %\vspace{-0.25cm}
  \caption{\TurtleKV{} WM Tuning Sensitivity}
  \label{figure:ycsb:vs_turtle}
  %\vspace{-0.25cm}
\end{figure*}

\vspace{-0.2cm}
\subsection{Experimental Setup}

\subsubsection{Hardware setup}  Experiments were run on an AMD Ryzen
Threadripper 7970x workstation (32 cores/64 threads), with 128 GB of
DDR5-6000 registered ECC memory, running Linux 6.3.  The SSD used was 
an Intel P4800x 1.5TB, with 512 byte LBA configured, formatted with XFS.  
The P4800x is a data center grade Optane device supporting PCIe 3.0, 
rated for2.4 GB/s read bandwidth and 2 GB/s write (550k/500k IOPs
respectively).

\vspace{-0.2cm}
\subsubsection{Software/Benchmark Configuration}

We use the Yahoo Cloud Serving Benchmark
(YCSB) \cite{cooperBenchmarkingCloudServing2010}, comparing \TurtleKV{}
to other state-of-the-art key-value engines, each representing a
different optimization target: WiredTiger, a read-optimized
\Bplustree{}-based engine; RocksDB, a write-optimized and widely used
\LSMtree{}-based system; and SplinterDB, a state-of-the-art
hybrid-optimized key-value store.  Each YCSB run inserts 400M records
comprised of 8 byte keys and 120 byte values (Load), then performs
150M operations comprised of 50\% updates/50\% gets (workload A),
another 150M operations comprised of 5\% updates/95\% gets (workload
B), 150M gets (workload C), 15M operations comprised of 95\% scans
(length $\le$ 100 keys)/5\% updates (workload E), and then a final 150M
operations comprised of 50\% read-modify-writes/50\% gets (workload
F). The total size of the initial data set is 48GB; to evaluate both
out-of-cache and in-cache performance, we run the benchmark twice for
each database, once with cache size 16GB ($\frac{1}{3}$ of data size)
and again with 48GB cache (large enough to fit the raw data).

\vspace{-0.2cm}
\subsubsection{Database Configurations} The configuration used for
each database is informed by the case studies described in
$\S$\ref{systemTuningConcerns}.  For WiredTiger, eviction dirty target
is set to half the cache size and eviction dirty trigger to 95\% of the
cache; transaction-level logging is enabled, and automatic checkpoints are disabled.
For RocksDB, parallelism and maximum parallel sub-compactions are set 
to the processor count (64); write buffer size is 64MB. RocksDB's block
cache is enabled (instead of the kernel's file cache) to accurately track memory usage; direct I/O is enabled, 
which we observe to be slightly faster. SplinterDB
is configured with mostly default values.  We set the maximum key size
to 8 bytes to match our test workload, and configure the number of
background threads to 2 for memtable tasks, 24 for normal tasks,
following the guidance of the SplinterDB tuning guide \cite{SplinterDBb}.

\TurtleKV{} is configured to use 32MB leaf pages, 20 bits per key for
per-leaf quotient filters (SplinterDB's default was left unchanged at
26).  Checkpoint distance was configured in two different ways: we
selected 4 different values for each cache size (128MB, 256MB, 512MB,
and 768MB for 16GB cache; 512MB, 1GB, 1.5GB, and 3GB for 48GB cache)
to run \TurtleKV{} \textit{statically tuned} (i.e., all YCSB workloads
run with the same setting).  We also selected, through trial and
error, a ``known good'' setting for checkpoint distance for each
workload and cache size combination: for 16GB cache,
Load$\rightarrow$512MB, A/B$\rightarrow$768MB,
C$\rightarrow$32MB, E$\rightarrow$192MB, and F$\rightarrow$640MB. For
48GB cache, Load/A/B$\rightarrow$3GB, C$\rightarrow$32MB,
E$\rightarrow$256MB, and F$\rightarrow$2GB.  We then configure the
test runner to run \TurtleKV{} \textit{dynamically tuned} by switching
to the ``known good'' value at the start of each workload.  Although
this is probably not a practical technique for real-world use, we
performed this test to verify the potential of \TurtleKV{}'s WM tuning
knob.  All comparisons with the other databases show \TurtleKV{}'s
dynamic tuning mode.

\vspace{-0.2cm}
\subsection{Results \& Discussion}\label{section:eval:results}

\subsubsection{YCSB Throughput: \TurtleKV{} vs Others}
Figure \ref{figure:ycsb:vs_world} shows the end-to-end throughput of
each YCSB workload, for dynamically-tuned \TurtleKV{} and all other
databases.

\TurtleKV{} performed excellently in the high-memory configuration
across all workloads (Figures \ref{figure:ycsb:vs_world:wr_48gb},
\ref{figure:ycsb:vs_world:rd_48gb}), exceeded only in
scan-dominated workload E by WiredTiger.  WiredTiger's \Bplustree{} demonstrates
a clear advantage for scans in both cache configurations, since
it does not require any in-memory merging, unlike \LSMtree{}s, \STBepsilontree{}s, and \TurtleTree{}s.

For cache-constrained write-heavy workloads (Figure
\ref{figure:ycsb:vs_world:wr_16gb}), SplinterDB was the highest
performer, which agrees with our case studies on WM tuning
($\S$\ref{systemTuningConcerns}).  In the cache-constrained read-heavy
workloads (Figure \ref{figure:ycsb:vs_world:rd_16gb}), \TurtleKV{} was
able to match or exceed the point query throughput of the other write-optimized
systems (RocksDB and SplinterDB), which is expected
because \TurtleTree{}'s level-tiered compaction policy is more
read-optimized than the lazy compaction of SplinterDB.  \TurtleKV{}'s
performance in workload E (scans) shows room for improvement; we
believe the low scan rate is due to cache pressure from the active
(unflushed/uncompacted) MemTable left over from previous workloads.
This data supports the idea that \TurtleKV{}'s scan performance in
cache-limited configurations could be improved by implementing a
mechanism to detect query-driven cache pressure and respond by
flushing the current MemTable to the checkpoint \TurtleTree{} to free
up memory.

\vspace{-0.25cm}
\subsubsection{Static vs Dynamic Tuning}
Figure \ref{figure:ycsb:vs_turtle} (a-d) shows \TurtleKV{}'s
sensitivity to checkpoint distance tuning parameter selection across
YCSB workloads.  Particularly noteworthy is the benefit of setting a
lower checkpoint distance for point-query dominated workloads B and C
(\ref{figure:ycsb:vs_turtle:rd_16gb},
\ref{figure:ycsb:vs_turtle:rd_48gb}).  The ability to give explicit
direction to the system as to how it should use memory helps
\TurtleKV{} to achieve much higher performance in these workloads than
the other systems (\ref{figure:ycsb:vs_world:rd_16gb},
\ref{figure:ycsb:vs_world:rd_48gb}), including \TurtleKV{} itself when
tuned statically.

\begin{figure*}
  \centering
  \vspace{-0.4cm}
  \subfloat[Insert, (16GB)]{{
    \includegraphics[width=0.67\columnwidth]{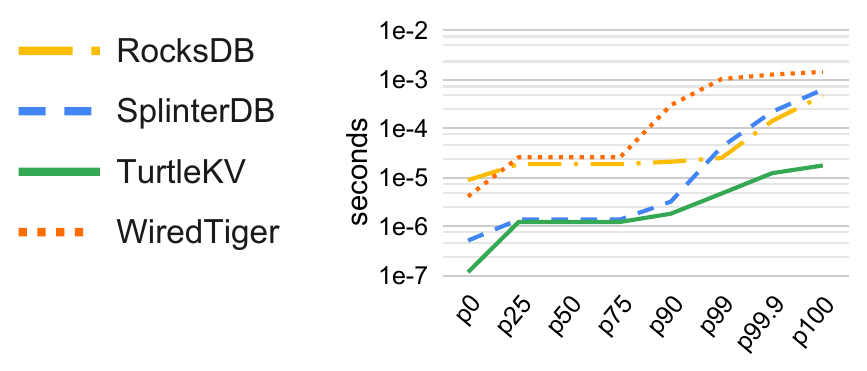}\label{figure:ycsb:latency:insert_16gb}
  \vspace{-0.2cm}
  }}
  \subfloat[Update (16GB)]{{
    \includegraphics[width=0.4\columnwidth]{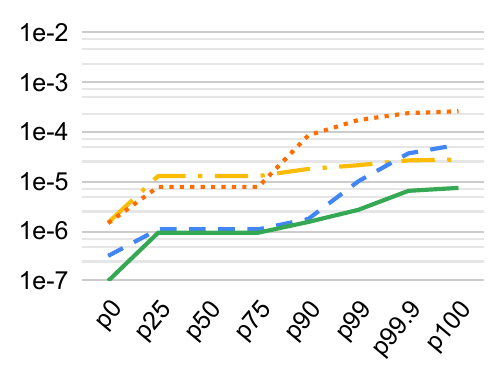}\label{figure:ycsb:latency:update_16gb}
  \vspace{-0.2cm}
  }}
  \subfloat[Get (16GB)]{{
    \includegraphics[width=0.4\columnwidth]{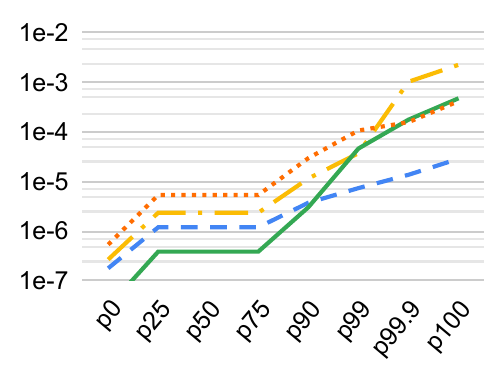}\label{figure:ycsb:latency:get_16gb}
  \vspace{-0.2cm}
  }}
  \subfloat[Scan (16GB)]{{
    \includegraphics[width=0.4\columnwidth]{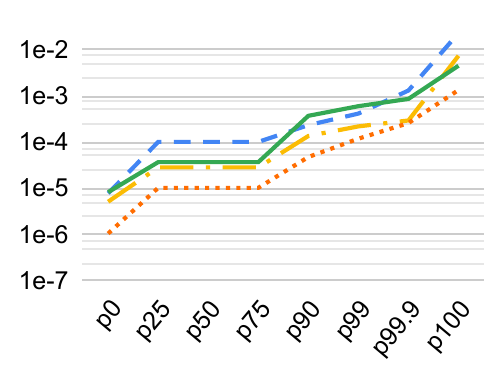}\label{figure:ycsb:latency:scan_16gb}
  \vspace{-0.2cm}
  }}
  \vspace{-0.4cm}
  \\   
  \subfloat[Insert (48GB)]{{
    \includegraphics[width=0.67\columnwidth]{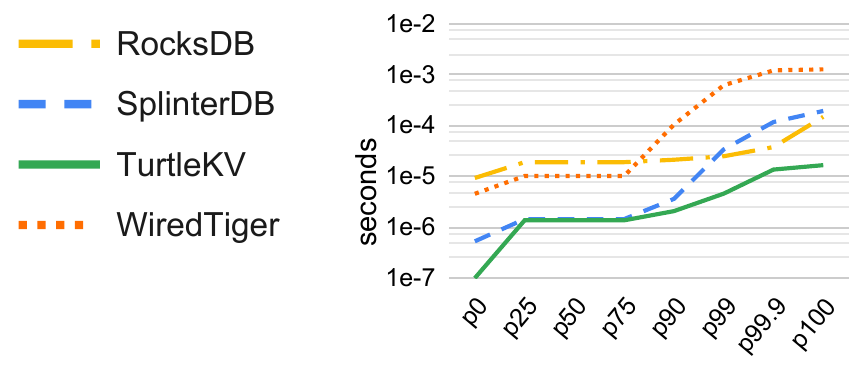}\label{figure:ycsb:latency:insert_48gb}
  \vspace{-0.2cm}
  }}
  \subfloat[Update (48GB)]{{
    \includegraphics[width=0.4\columnwidth]{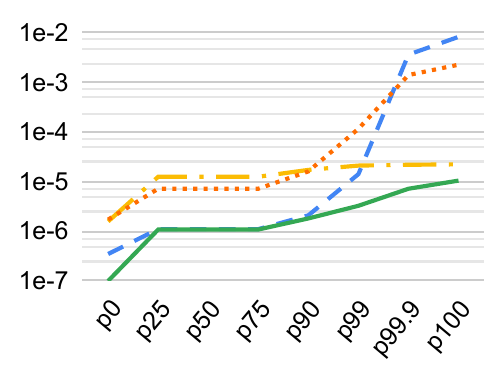}\label{figure:ycsb:latency:update_48gb}
  \vspace{-0.2cm}
  }}
  \subfloat[Get (48GB)]{{
    \includegraphics[width=0.4\columnwidth]{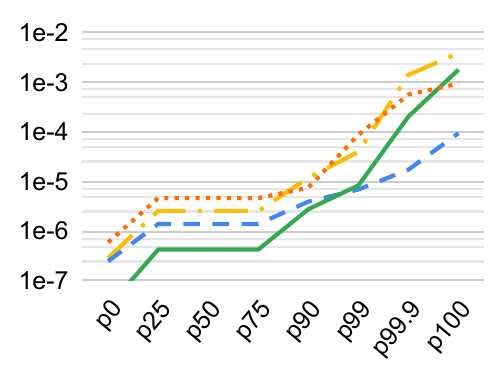}\label{figure:ycsb:latency:get_48gb}
  \vspace{-0.2cm}
  }}
  \subfloat[Scan (48GB)]{{
    \includegraphics[width=0.4\columnwidth]{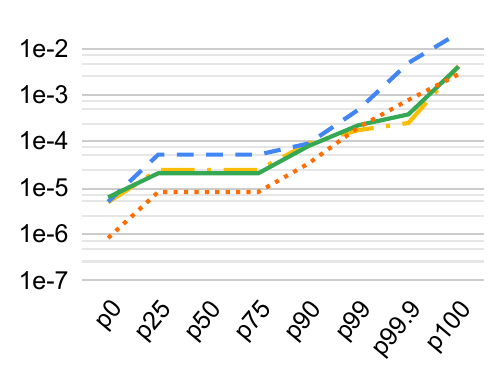}\label{figure:ycsb:latency:scan_48gb}
  \vspace{-0.2cm}
  }} 
  \vspace{-0.25cm}
  \caption{YCSB, Per-Operation Latency Distributions for 16GB and 48GB Cache Sizes}
  \label{figure:ycsb:latency}
  \vspace{-0.5cm}
\end{figure*}

Figure \ref{figure:ycsb:vs_turtle:scale_invariant} demonstrates our
claim that the write-optimization benefit of WM tuning in
\TurtleTree{}s scales independently of total dictionary size $N$.  The
higher variation in the smallest dataset (N=31M) is because this
scenario is more sensitive to small differences in the number of
batches flushed down the tree; the leaf page size $L$ is configured to
32MB, which is a larger percentage of the overall data size at smaller
scale.

\vspace{-0.1cm}
\subsubsection{YCSB Latency Percentiles}
Figure \ref{figure:ycsb:latency} shows the per-operation latency
distributions for the YCSB tests.  A key result in this data is that
\TurtleKV{} demonstrates superior tail latencies (P99.9) for both
inserts and updates, supporting the effectiveness of \TurtleKV{}'s
update pipeline and multi-processor approach to checkpoint updates.
\TurtleKV{} falls behind SplinterDB in low-cache, insert-heavy workloads due to high P100
during checkpoint pipeline stalls.

While latencies for queries are generally good, \TurtleKV{} does
present somewhat elevated tail latencies for point queries with both high and low
cache sizes, and for small-cache scans.\iffalse{}, suggesting possible room for
improvement in \TurtleTree{} search and I/O (caching).\fi{}

\begin{figure}
\vspace{-0.15cm}
  \centering
  \subfloat[48GB Cache]{{
    \includegraphics[width=0.85 \columnwidth]{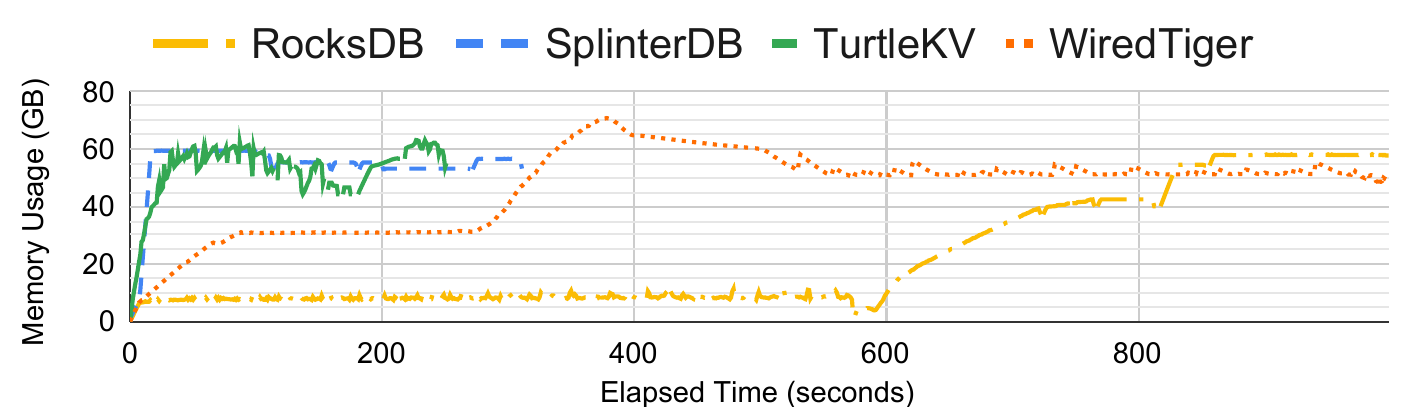}\label{figure:ycsb:memory_48gb}
  }}
  \vspace{-0.25cm}
  \subfloat[16GB Cache]{{ \includegraphics[width=0.85
        \columnwidth]{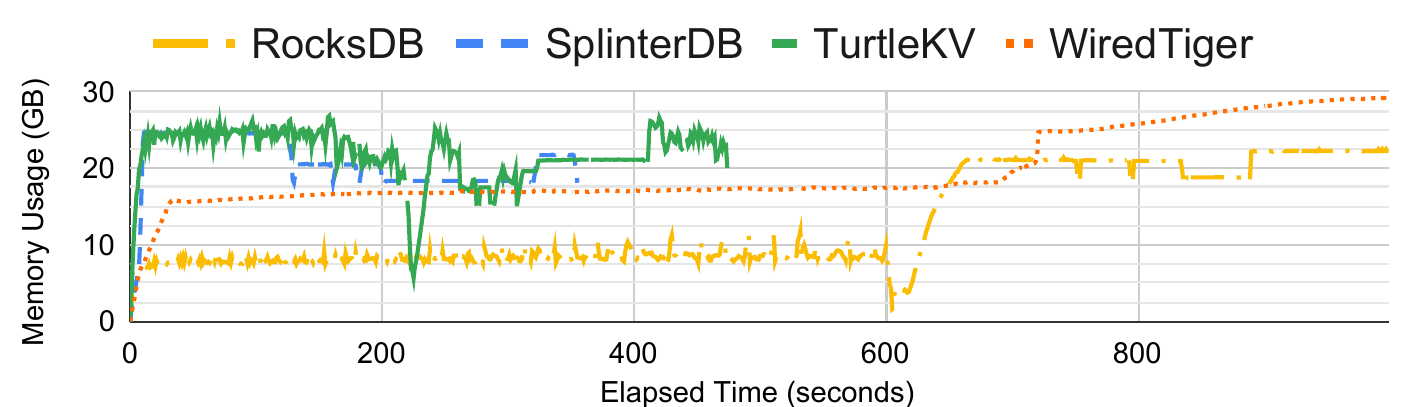}\label{figure:ycsb:memory_16gb}
  }} \\
  \vspace{-0.25cm}
  \caption{Memory Usage over Time During YCSB}
  \label{figure:ycsb:memory}
  \vspace{-0.5cm}
\end{figure}

\vspace{-0.1cm}
\subsubsection{Memory Usage over Time}
Figure \ref{figure:ycsb:memory} shows memory usage in the benchmark
process for each database over the course of the YCSB workloads.
\TurtleKV{} and SplinterDB both quickly make use of available memory
during the Load; afterwards, both generally stayed below their
allocated limit.  Note, because the load generator is in the same
process as the key-value stores being tested, and because we used
pre-recorded workload trace files that are buffered in memory, the
overall memory footprint is somewhat higher than the configured cache
limit.  Both RocksDB and WiredTiger did not make use of all available
headroom during write-heavy workloads.  RocksDB did slightly better
than \TurtleKV{} and SplinterDB at staying within the configured
limit.  WiredTiger presented the highest memory overhead during 
the benchmark.

\begin{figure}
  \centering
    \includegraphics[width=0.75 \columnwidth]{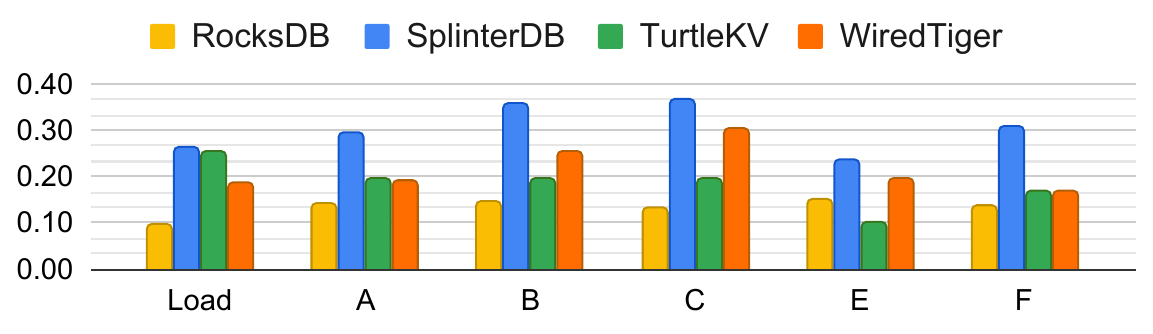}
  \vspace{-0.35cm}
  \caption{CPU Cache Miss Rate, 48GB cache}
  \label{figure:ycsb:cpu_cache_48gb}
  \vspace{-0.4cm}
\end{figure}

\vspace{-0.2cm}
\subsubsection{CPU Efficiency for In-Cache}
Figure \ref{figure:ycsb:cpu_cache_48gb} shows the CPU cache miss rate
for each workload of YCSB.  \TurtleKV{} is particularly cache-friendly
in query-heavy workloads C and E, due to its careful attention to page
layout (cache-oblivious trie indexes at the head of each leaf page)
and use of ARTs.  In mixed workloads A and F, \TurtleKV{} performs
similarly in this metric to leaders RocksDB and WiredTiger.  For the
Load workload, CPU cache miss rate is elevated due to the shuffling
effect of data-parallel batch compactions while updating the
\TurtleTree{} checkpoint; this is comparable with SplinterDB's cache miss
rate for the Load phase.

%% file: sections/6-related-work.tex
%\vspace{-0.15cm}
\section{Related Work}

The theoretical trade-off space constraining external memory indexing
is characterized by Brodal and Fagerberg
\cite{brodalLowerBoundsExternal2003} for comparison-based
dictionaries, and by Conway et
al. \cite{conwayOptimalHashingExternal2018} for hash-based
dictionaries.  The key insights of these works are echoed and
elaborated in practical terms by the RUM Conjecture
\cite{athanassoulisDesigningAccessMethods2016}.

Recent efforts to explore trade-offs in LSM-tree based systems include
Monkey \cite{dayanMonkeyOptimalNavigable2017}, a key-value store which
optimizes queries through an analytically optimal allocation of memory
to Bloom filters at various levels of the tree; Dostoevsky
\cite{dayanDostoevskyBetterSpaceTime2018}, which further improves
space-time trade-offs for LSM-based systems by varying compaction
policy from lazy at lower levels to greedy at higher ones; Wacky
\cite{dayanLogStructuredMergeBushWacky2019}, which introduces adaptive
level sizing and finer-grained control over compaction policy to even
better optimize the trade-off curve; and Cosine
\cite{chatterjeeCosineCloudcostOptimized2021}, which builds on the
previous three and uses machine learning to optimize data structure
design for specific cloud-based key-value storage workloads.  We note
that Dostoevsky and Wacky in particular share some thematic similarity
to \TurtleKV{}'s scheme of optimizing writes without harming future
query performance by deferring on-disk compaction of recent updates
and compacting older data more aggressively. Endure
\cite{huynhvldb2022} tunes size ratio, memory allocations, and
compaction policy to maximize throughput under workload uncertainty.

Notable work in extending the trade-off tuning potential of B-tree
based stores includes the Bw-Tree \cite{levandoskiBwTreeBtreeNew2013,
  wangBuildingBwTreeTakes2018}, a log-structured B-Tree which uses
lock-free techniques to achieve low overheads for in-memory operations
in high concurrency settings; Bf-Trees, a similar approach which
introduces the innovation of caching \textit{mini-pages} using a
circular buffer to reduce I/O amplification while minimizing the
overhead of cache maintenance; and the Bw$^e$-Tree
\cite{wangBweTreeEvolution2024}, which describes many practical
improvements to the Bw-Tree design to optimize for fast SSDs.
LSB-Trees \cite{kimLSBTreeLogstructuredBTree2015} propose an alternate
mechanism for storing B-Tree deltas using a combination of physical
and logical logging.  The relationship of Log-Structured B-Trees to
the overall continuum of trade-offs in the external memory dictionary
design space is explored by Idreos et
al. \cite{idreosDesignContinuumsPath2019}.

Several recent works address key-value engine design in the face of
very fast modern storage devices.  SplinterDB
\cite{conwaySplinterDBClosingBandwidth2020} and Mapped SplinterDB
\cite{conwaySplinterDBMapletsImproving2023} introduce innovations in
the design of external memory dictionaries, extending
B$^\epsilon$-Trees to STB$^\epsilon$-Trees by applying ideas from
size-tiered LSM-trees to achieve high update efficiency without
excessively penalizing point queries.  FASTER
\cite{chandramouliFASTERConcurrentKeyValue2018,
  chandramouliFASTEREmbeddedConcurrent2018} combines log-structured
storage with in-memory concurrent hash tables to scale
point-query-only based workloads to modern SSDs, but lacks first-class
support for range operations.  KVell
\cite{lepersKVellDesignImplementation2019a}, later extended to KVell+
\cite{lepersKVellSnapshotIsolation2020}, achieves high performance on
modern SSDs using a \textit{shared-nothing} design based on
independent sharded indices maintained on separate threads.  KVell
only maintains key-based indices in memory (not in storage), trading
scalability and recovery time for lower space-/write-amplification.

%% file: sections/7-conclusion.tex
\section{Conclusion \& Future Work}

With \TurtleKV{}, we introduce a novel approach to read and write
optimization that dynamically trades memory to establish a desired
trade-off point \textit{on demand}. The \TurtleTree{} design and its
accompanying tuning knob, checkpoint distance ($\chi$), achieves this
property by creating an unbiased on-disk structure whose read and
write I/O cost can be directly tuned via dynamic memory allocation
without requiring external data structures that permanently consume
available memory resources.

This dynamic approach to performance optimization opens many paths for
future research and improvement of key-value storage techniques. These
include extending the dynamic range of $\chi$-based tuning by further
optimizing in-memory checkpoint tree updates, reducing memory
requirements through the use of streaming compaction and other
techniques, and exploring more of the B$^{\epsilon+}$-Tree design
space to discover data structures beyond the \TurtleTree{}.